\documentclass[prd,superscriptaddress,nofootinbib,preprintnumbers,amsmath,amssymb,10pt,twocolumn]{revtex4-1}
\usepackage[utf8]{inputenc}
\usepackage{amsfonts}
\usepackage{mathrsfs}
\usepackage{physics}
\usepackage{footnote}
\usepackage{scrextend}
\usepackage{leftidx}
\usepackage{soul}
\usepackage{braket}
\usepackage{makecell}
\usepackage{amsmath,amsbsy}
 \usepackage{array}
 \newcolumntype{C}[1]{>{\centering\arraybackslash}p{#1}}
 
\usepackage[dvipsnames]{xcolor}
\definecolor{red}{rgb}{0.9, 0,0}
\definecolor{cerulean}{rgb}{0., 0.42,0.9}

\usepackage{multirow}
\usepackage{epsfig}
\usepackage{graphicx}  
\usepackage{url}
\usepackage{color}
\usepackage{hyperref}
\hypersetup{
     colorlinks   = true,
     citecolor    = red,
	 linkcolor=blue
}
\usepackage{float}
\usepackage[export]{adjustbox}
\usepackage{multirow}
\usepackage[lofdepth,lotdepth,caption=false,justification=RaggedRight]{subfig}

\definecolor{red}{rgb}{0.9, 0,0}
\definecolor{cerulean}{rgb}{0., 0.62,0.9}
\definecolor{navy}{rgb}{0.05, 0.05,0.8}

\graphicspath{{./figures/}}

\newcommand{\bfv}{{\bf v}}

\newcommand{\nn}{\nonumber}

\usepackage[dvipsnames]{xcolor}

\begin{document}

\title{Anharmonic effects in nuclear recoils from sub-GeV dark matter}
\author{Tongyan Lin}
\affiliation{ Department of Physics, University of California, San Diego, CA 92093, USA }
\author{Chia-Hsien Shen}
\affiliation{ Department of Physics, University of California, San Diego, CA 92093, USA }
\affiliation{Department of Physics and Center for Theoretical Physics,
National Taiwan University, Taipei 10617, Taiwan}
\author{Mukul Sholapurkar}
\affiliation{ Department of Physics, University of California, San Diego, CA 92093, USA }
\author{Ethan Villarama}
\affiliation{ Department of Physics, University of California, San Diego, CA 92093, USA }

\date{\today}

\begin{abstract}\noindent Direct detection experiments are looking for nuclear recoils from scattering of sub-GeV dark matter (DM) in crystals, and have thresholds as low as $\sim 10~\text{eV}$ or DM masses of $\sim 100~\text{MeV}$. Future experiments are aiming for even lower thresholds. At such low energies, the free nuclear recoil prescription breaks down, and the relevant final states are phonons in the crystal. Scattering rates into single as well as multiple phonons have already been computed for a harmonic crystal. However, crystals typically exhibit some anharmonicity, which can significantly impact scattering rates in certain kinematic regimes. In this work, we estimate the impact of anharmonic effects on scattering rates for DM in the mass range $\sim 1-10$ MeV, where the details of multiphonon production are most important. Using a simple model of a nucleus in a bound potential, we find that anharmonicity can modify the scattering rates by up to two orders of magnitude for DM masses of $\mathcal{O}(\text{MeV})$. However, such effects are primarily present at high energies where the rates are suppressed, and thus only relevant for very large DM cross sections. We show that anharmonic effects are negligible for masses larger than $\sim 10~\text{MeV}$. 
\end{abstract}

\maketitle

\tableofcontents
\section{Introduction}
\label{sec:intro}

Over the past few decades, a significant theoretical and experimental effort has been dedicated to detect dark matter (DM), but the particle nature of DM still remains a mystery. Direct detection experiments look for the direct signatures left by halo DM depositing energy inside the detectors. Traditionally, such experiments have looked for elastic nuclear recoils induced by DM particles in detectors~\cite{Goodman:1984dc}. This strategy has had tremendous sensitivity for DM particles with masses higher than the GeV-scale that interact with nuclei~\cite{Aprile_20232,LZ:2022ufs1,PandaX-4T:2021bab}. However, in recent years it has also been recognized that sub-GeV dark matter models are also compelling and motivated dark matter candidates~\cite{Boehm_2004,B_hm_2004,Fayet_2004,Pospelov:2007mp,Feng:2008ya,Kaplan:2009ag,Essig:2011nj}. These DM particles would leave much lower energy nuclear recoils, motivating experimental efforts to lower the detector thresholds for nuclear recoils. Inelastic processes like the Migdal effect~\cite{Migdal1939,Ibe:2017yqa,Essig:2019xkx,Knapen:2020aky,Berghaus:2022pbu} or bremsstrahlung~\cite{Kouvaris:2016afs} provide alternative channels to detect nuclear scattering in the sub-GeV DM regime. 

The majority of experiments achieving lower thresholds in nuclear recoils (down to $\sim$ 10 eV) are doing so with crystal targets~\cite{Alkhatib_20211,Angloher_2023,Abdelhameed_2019,DAMIC:2020cut}, although there is also progress in using liquid helium~\cite{Anthony-Petersen:2023ykl}. Future experiments like SPICE~\cite{SPICE7} will reach even lower thresholds by measuring athermal phonons produced in crystals like GaAs and Sapphire (i.e. $\text{Al}_2\text{O}_3$). In crystal targets, DM-nucleus scattering can deviate substantially from the picture of a free nucleus undergoing elastic recoils. Nuclei (or atoms) are subject to forces from the rest of the lattice, which play a role at the lower energies relevant for sub-GeV DM. For recoil energies below the typical binding energy of the atom to the lattice ($\mathcal{O}$(10~\text{eV})), the atoms are instead treated as being bound in a potential well. At even lower energies, the relevant degrees of freedom are the collective excitations of the lattice, known as phonons. In this regime, single phonon excitations with typical energies $\lesssim 0.1$ eV are possible. 

In the DM scattering rate, crystal scattering effects are all encoded within a quantity known as the dynamic structure factor, $S(\textbf{q},\omega)$. The differential cross section for a DM particle of velocity $\textbf{v}$ and mass $m_\chi$ to scatter with energy deposition $\omega$ and momentum transfer $\textbf{q}$ can be written in terms of  $S(\textbf{q},\omega)$ as:
\begin{align} 
\label{eq:differential_cross_sec}
\frac{d\sigma}{d^3\textbf{q}d \omega} = \frac{b_p^2}{\mu_\chi^2} \frac{1}{v}\frac{\Omega_c}{2\pi} |\Tilde{F}(\textbf{q})|^2 S(\textbf{q},\omega) \delta (\omega - \omega_{\textbf{q}}), 
\end{align}
Here $b_p$ is the scattering length of the DM with a proton, $\mu_\chi$ is the reduced DM-proton mass, $\Omega_c \equiv V/N$ is the volume of the unit cell in the crystal with total volume $V$ and $N$ unit cells, and $\omega_{\textbf{q}} = \textbf{q} \cdot \textbf{v} - q^2/2m_\chi$ is equal to the energy $\omega$ lost by the DM particle when it transfers momentum $\textbf{q}$ to the lattice. The $\textbf{q}$-dependence of the DM-nucleus interaction is encapsulated in the DM form factor $\Tilde{F}(\textbf{q})$. $S(\textbf{q},\omega)$ can thus be viewed as a form factor for the crystal response. For a recent review, see Ref.~\cite{Kahn:2021ttr}.

Understanding $S(\textbf{q}, \omega)$ in crystals is critical to direct detection of sub-GeV dark matter.  Thus far, the limiting behavior of  $S(\textbf{q}, \omega)$ is well understood~\cite{Campbell-Deem:2022fqm}. In the limit of large $\omega$ and $q$ ($\omega \gtrsim$ eV and $q \sim \sqrt{ 2 m_N \omega}$ for nucleus of mass $m_N$), the structure factor behaves as $S(\textbf{q}, \omega) \propto \delta\left(q^2/(2 m_N) - \omega \right)$, reproducing the cross section for free elastic recoils. At low $\omega$ comparable to the typical phonon energy $\omega_0$ and $q$ comparable to the inverse lattice spacing, $S(\textbf{q}, \omega)$ instead is dominated by single phonon production. The intermediate regime, particularly $q \sim \sqrt{2 m_N \omega_0}$, is dominated by multiphonon production. For a large number of phonons being produced, this should merge into the free nuclear recoil limit.

For DM masses below $\sim$MeV, the momentum-transfers are smaller than the typical inverse lattice spacing of crystals, $q < 2\pi/a \sim {\mathcal{O}}({\rm keV})$,  where $a$ is the lattice spacing. The dominant process is the production of a single phonon. In recent years, the single phonon contribution to $S(\textbf{q}, \omega)$ has been computed extensively in a variety of materials, often using first-principles approaches for the phonons~\cite{Knapen:2017ekk,Griffin:2018bjn,Trickle:2019nya,Cox:2019cod,Griffin:2019mvc,Mitridate:2020kly,Trickle:2020oki,Griffin:2020lgd,Coskuner:2021qxo,Knapen:2021bwg,Mitridate:2023izi}. In most of the crystals, single phonons have a maximum energy of $\mathcal{O}(100~\text{meV})$, however,  requiring extremely low experimental thresholds to detect them.

\begin{figure*}[t]
	\centering
 \begin{minipage}{7in}
  \centering
  \raisebox{-0.5\height}{\includegraphics[height=3.3in]{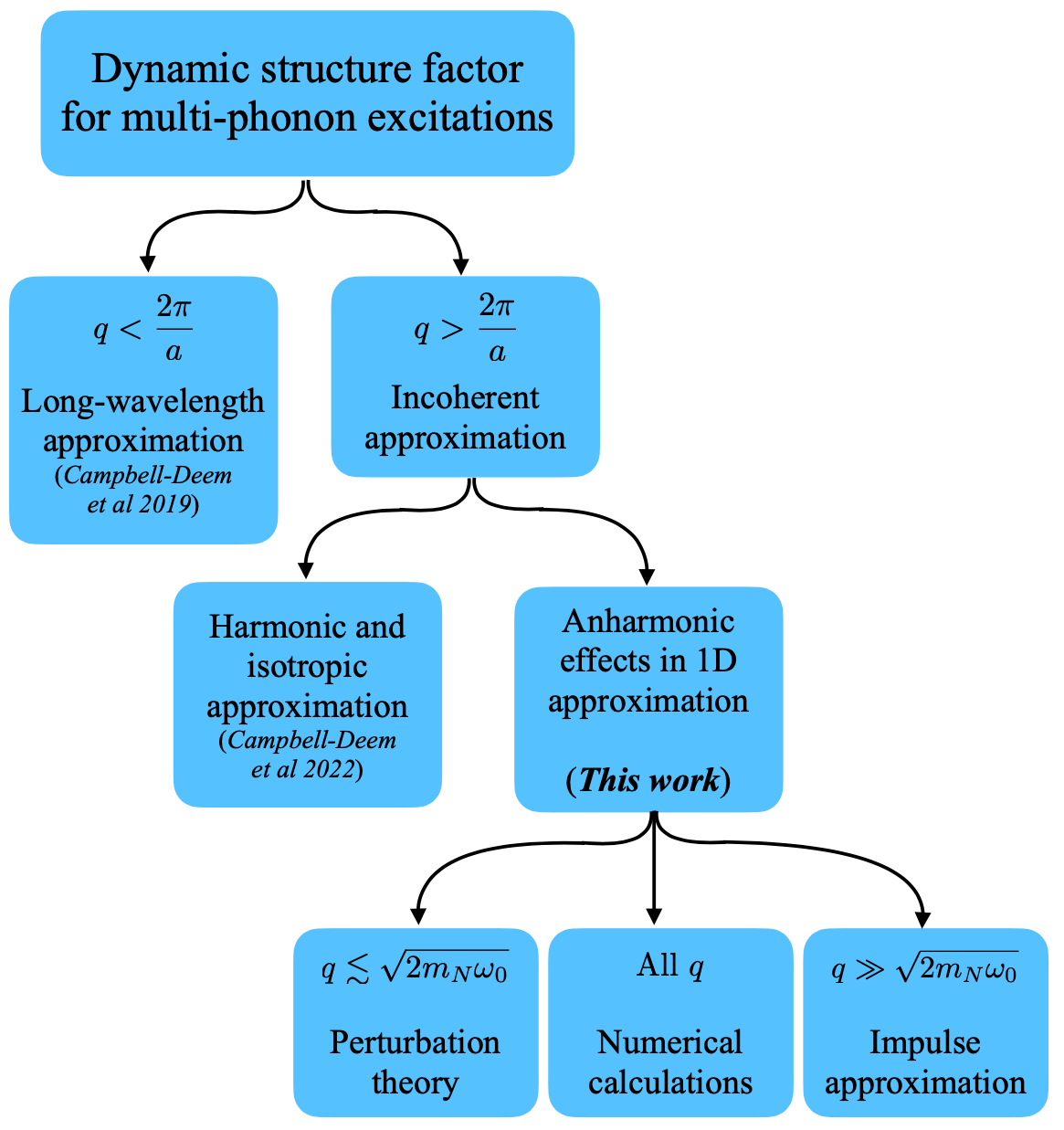}}
  \hspace*{.2in}
  \raisebox{-0.5\height}{\includegraphics[height=2.6in]{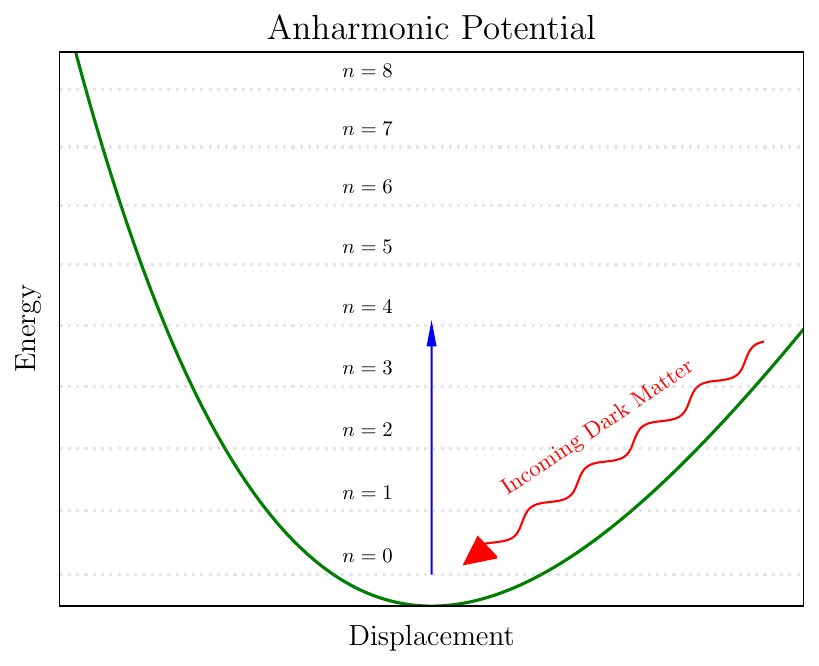}}
\end{minipage}
	\caption{({\bf Left}) Due to the computational challenges of obtaining the multiphonon scattering rate in crystals, analytic approximations are valuable. Here we show a classification of regimes in which a multiphonon calculation has been performed, as well as approximations made in each case. In this work, we show that anharmonic corrections can be significant for $q \lesssim \sqrt{2 m_N \omega_0}$ (Sec.~\ref{sec:perturbation_theory}) but are negligible when $q \gg \sqrt{2 m_N \omega_0}$ (Sec.~\ref{sec:impulse}). We obtain results for all $q$ using numerical calculations (Sec.~\ref{sec:scattering_rates}). ({\bf Right}) To estimate anharmonic effects, we take a toy model of dark matter scattering off an atom in a 1D anharmonic potential. We obtain the anharmonicity by fitting to empirical models of interatomic potentials. }
 \label{fig:flowchart}
\end{figure*}

Production of multiphonons is an enticing channel to look for sub-GeV DM with detectors having thresholds higher than $\mathcal{O}(100~\text{meV})$. They are also important to understand in the near term as experiments lower their thresholds. However, multiphonon production has been more challenging to compute. The numerical first-principles approach taken for single phonon production does not scale well with number of phonons being produced, where even the two-phonon rate becomes very complicated. Alternate analytic methods are thus valuable. In Fig.~\ref{fig:flowchart}, we show a classification of the different regimes in which a multiphonon calculation has been performed, including this work. We discuss the details of these regimes and calculations below.

One analytic approach was taken in Ref.~\cite{Campbell-Deem:2019hdx}, which calculated the two-phonon rate in the long-wavelength limit, but this study was limited to the regime $q < 2\pi/a$ and focused on acoustic phonons only. For $q > 2\pi/a$, a different approximation is possible, the incoherent approximation, which drops interference terms between different atoms of the crystal in calculating $S(\textbf{q}, \omega)$. Then scattering is dominated by recoiling off of individual atoms. This approach was taken in~\cite{Campbell-Deem:2022fqm}, which found a general $n$-phonon production rate scaling as $(q^2/(2m_N \omega_0))^n$. This result also showed how the free nuclear recoil cross section was reproduced in the multiphonon structure factor as $ q \gg \sqrt{2 m_N \omega_0}$. 

However, one limitation of the multiphonon production rate in Ref.~\cite{Campbell-Deem:2022fqm} was that it worked in the harmonic approximation, where higher order phonon interactions like the three-phonon interaction are neglected. Typical crystals have some anharmonicity which introduces phonon self-interactions, leading to various observable effects like phonon decays, thermal expansion, and thermal conductivity of crystals~\cite{srivastava2019physics,DEBERNARDI19991,Wei_2021}. Using a simplified model of anharmonic phonon interactions, Ref.~\cite{Campbell-Deem:2022fqm} estimated that anharmonic three-phonon interactions may give the dominant contribution to the two-phonon rate $q < 2 \pi/a$, and are larger than the harmonic piece by almost an order of magnitude in the regime. On the other hand, we do not expect anharmonic effects to be important in the opposite limit of large $q$ ($q \gg \sqrt{2m_N \omega_0}$), where the nucleus can be treated as a free particle. It is thus necessary to bridge these two extremes and estimate the anharmonic effects in the intermediate regime where multi-phonons dominate the scattering.

In this work, we estimate the anharmonic effects on the rate of multiphonon production by working in the incoherent approximation and $q > 2\pi/a$. In this limit, the multiphonon scattering rate looks similar to that of an atom in a potential~\cite{Kahn:2020fef}, although the spectrum of states is smeared out due to interactions between neighboring atoms. Given this similarity, we will take a toy model of an atom in a 1D potential. This gives a simple approach to including anharmonic effects, which is also illustrated in the right panel of Fig.~\ref{fig:flowchart}. The anharmonic corrections to the atomic potential only capture a part of the contributions to anharmonic phonon interactions, but they have a similar size (in the appropriate dimensionless units) and should give a reasonable estimate of the size of the effect. We can therefore use this approach to estimate theoretical uncertainties and gain analytic understanding for the multiphonon production rate. However, the result should not be taken as a definitive calculation of the anharmonic corrections. Fortunately, we will find that anharmonic corrections are large only in certain parts of the phase space which are more challenging to observe, and that the multiphonon rate quickly converges to the harmonic result for DM masses above a few MeV.

The outline of this paper is as follows: In Sec.~\ref{sec:scattering}, we discuss the formalism of DM scattering in a crystal and the dynamic structure factor, which encodes the information about the crystal response. We consider the calculation of the structure factor under the incoherent approximation, and motivate the anharmonic 1D toy potentials we use in this paper. In Sec.~\ref{sec:analytic}, we study the behavior of the dynamic structure factor analytically for the anharmonic 1D potentials. Using perturbation theory, we show that anharmonic corrections can dominate for $q \ll \sqrt{ 2 m_N \omega_0}$ and become more important for higher phonon number. In the opposite limit $q \gg \sqrt{ 2 m_N \omega_0}$, we use the impulse approximation to show that anharmonic corrections are negligible and that the structure factor indeed approaches that of an elastic recoil. In Sec.~\ref{sec:exact}, we present numerical results for the structure factor in anharmonic 1D potentials obtained from realistic atomic potentials in various crystals. 
In Sec.~\ref{sec:scattering_rates}, we calculate the impacts of including anharmonic effects on DM scattering rates. We conclude in Sec.~\ref{sec:conclusions}. 

Appendix~\ref{appendix:interatomic_potentials} gives the details of the modeling of the interatomic forces on the lattice, used to extract 1D single atom potentials. Appendix~\ref{appendix:perturbation_theory} gives additional details of the analytic perturbation theory estimates of the anharmonic structure factor. Appendix~\ref{sec:impulse_appendix} includes additional details relevant to the impulse approximation calculation. Appendix~\ref{sec:morse} summarizes the exactly solveable Morse potential model, which further validates the results in the main text.

\section{Dark matter scattering in a crystal}
\label{sec:scattering}

Consider DM that interacts with nuclei in the crystal. We will parameterize the interaction with the lattice by a coupling strength $f_{\boldsymbol{\ell} d}$ relative to that of a single proton, where $\boldsymbol{\ell}$ denotes the lattice vector of a unit cell and $d$ denotes the atoms in the unit cell. In the DM scattering cross section, \eqref{eq:differential_cross_sec}, the material properties of the crystal are encoded in the structure factor $S(\textbf{q},\omega)$ which is defined as,
\begin{align}
\label{eq:structurefactor}
    S(\textbf{q},\omega) \equiv \frac{2\pi}{V}\sum_f & \Big|\sum_{\boldsymbol{\ell}}\sum_{d}f_{\boldsymbol{\ell} d}\langle \Phi_f|e^{i\textbf{q} \cdot \textbf{r}_{\boldsymbol{\ell}d}}|0\rangle\Big|^2\nonumber\\& \times \delta(E_f - E_0 -\omega),
\end{align}
where $|\Phi_f\rangle$ is the final excited state of the crystal with energy $E_f$ and $\textbf{r}_{\boldsymbol{\ell}d}$ denotes the position of the scattered nucleus. The crystal is considered to be in the ground state $|0\rangle$ initially. Note for simplicity we assume a pure crystal where each atom has a unique coupling strength; the scattering is modified if there is a statistical distribution for the interaction strengths at each lattice site, for instance if different isotopes are present~\cite{Campbell-Deem:2022fqm}. 

The states $|\Phi_f\rangle$ are the phonon eigenstates of the lattice Hamiltonian,
\begin{align}
    H_{\text{lattice}}= \sum_{\boldsymbol{\ell} d} \frac{p_{\boldsymbol{\ell}d}^2}{2m_{\boldsymbol{\ell}d}} + V_{\text{lattice}} + E_0,
\end{align}
where the first term is the kinetic energy of the atoms in the lattice and the lattice potential $V_{\text{lattice}}$ in general is given by,
\begin{align}
    \label{eq:latticepotential}
    V_{\text{lattice}} &=  \frac{1}{2}\sum_{\boldsymbol{\ell},d,\boldsymbol{\ell}',d'} \sum_{\alpha,\beta}k_{\alpha\beta}^{(2)}(\boldsymbol{\ell}d,\boldsymbol{\ell}'d')~ u_{\alpha}(\boldsymbol{\ell}d) ~u_{\beta}(\boldsymbol{\ell}'d')\nonumber\\& + \frac{1}{3!}\sum_{\boldsymbol{\ell},d,\boldsymbol{\ell}',d',\boldsymbol{\ell}'',d''}\sum_{\alpha,\beta,\gamma} k_{\alpha\beta\gamma}^{(3)}(\boldsymbol{\ell}d,\boldsymbol{\ell}'d',\boldsymbol{\ell}''d'') \nonumber\\& \qquad \qquad \qquad \quad \times u_{\alpha}(\boldsymbol{\ell}d) ~u_{\beta}(\boldsymbol{\ell}'d')~u_{\gamma}(\boldsymbol{\ell}''d'')\nonumber\\&+...
\end{align}
where the $u_{\alpha}(\boldsymbol{\ell}d)$ is the displacement from the equilibrium position in the Cartesian direction $\alpha$ for the atom at the position $d$ in the unit cell located at $\boldsymbol{\ell}$, and  $k_{\alpha\beta}^{(2)}$, $k_{\alpha\beta\gamma}^{(3)}$ are the second-, and third-order force constants respectively. Note that as the displacements are considered around equilibrium, we do not have a term in the potential which is linear in the displacements. 

A number of approximations are useful in evaluating $S({\bf q}, \omega)$. The first is the harmonic approximation, which amounts to keeping the terms up to second-order force constants and neglecting the higher order terms ($k^{(3)}_{\alpha \beta \gamma} = 0)$. This vastly simplifies the Hamiltonian into a harmonic oscillator system, and has been used in most previous calculations of DM scattering in crystals. While this is generally an excellent approximation in crystals, including higher order terms in the Hamiltonian (anharmonicity) is necessary to explain a number of observable effects, as we will discuss further below.

The second approximation is the incoherent approximation, used for scattering with momentum transfers much bigger than the inverse lattice spacing of the crystal, $q \gg 2\pi/a$. In this limit, we drop the interference terms between different atoms in the crystal in \eqref{eq:structurefactor}. This amounts to summing over the squared matrix elements of individual atoms in the structure factor in \eqref{eq:structurefactor},
\begin{align}
    \label{eq:structure_factor_inc}
    S(\textbf{q},\omega) &\approx \frac{2\pi}{V}\sum_f \sum_{\boldsymbol{\ell}}\sum_{d} |f_{\boldsymbol{\ell} d}|^2\Big|\langle \Phi_f|e^{i\textbf{q} \cdot \textbf{r}_{\boldsymbol{\ell}d}}|0\rangle\Big|^2 \nonumber \\ & \times \delta(E_f - E_0 -\omega).
\end{align}
The calculation of the structure factor then simplifies to computing matrix elements $\Big|\langle \Phi_f|e^{i\textbf{q} \cdot \textbf{r}_{\boldsymbol{\ell}d}}|0\rangle\Big|^2$ which are identical for the atoms in all unit cells $\ell$. 

Below, we will first discuss this calculation under the approximation of a harmonic crystal, before going on to setting up a model that accounts for anharmonicity in crystals.

\subsection{Harmonic approximation}
\label{sec:structure_factor_harmonicToToy}

In the harmonic approximation, the lattice Hamiltonian can be written as a sum of harmonic oscillators in Fourier space~\cite{Schober2014},
\begin{align}\label{eq:harmonic_hamiltonian}
H_{\text{lattice}}^{\text{Harmonic}} = \sum_{\nu}^{3n}\sum_\textbf{q}\omega_{\textbf{q},\nu}(\hat{a}_{\textbf{q},\nu}^{\dagger}\hat{a}_{\textbf{q},\nu} + \frac{1}{2}),
\end{align}
where the phonon eigenmodes of the lattice are labelled by the momentum $\textbf{q}$ and the $3n$ branches $\nu$ with $n$ being the number of atoms in the unit cell. The $\hat{a}_{\textbf{q},\nu}^{\dagger}$   ($\hat{a}_{\textbf{q},\nu}$) are the creation (annihilation) operators, and $\omega_{\textbf{q},\nu}$ are the energies of the phonons. The lattice eigenstates that appear in \eqref{eq:structurefactor} can then be written as,
\begin{align}
|\Phi_n\rangle = \hat{a}_{\textbf{q}_1,\nu_1}^{\dagger} \hat{a}_{\textbf{q}_2,\nu_2}^{\dagger}...\hat{a}_{\textbf{q}_n,\nu_n}^{\dagger}|0\rangle,
\end{align}
where $|\Phi_n \rangle$ is an $n$-phonon state. The displacement operators in this harmonic approximation are given by,
\begin{align}\label{eq:displacementoperators}
\textbf{u}(\boldsymbol{\ell}d) = \sum_{\nu}^{3n}\sum_\textbf{q}\sqrt{\frac{1}{2Nm_d\omega_{\textbf{q},\nu}}}\Big( & \textbf{e}_{\textbf{q},\nu}(d)\,\hat{a}_{\textbf{q},\nu}\, e^{i\textbf{q}.\textbf{r}_{\boldsymbol{\ell}d}^0 - i\omega_{\textbf{q},\nu}t} \nonumber\\ & +~\text{h.c.} \Big),
\end{align}
where the $\textbf{e}_{\textbf{q},\nu}(d)$ indicates the eigenvector of the displacement of atom $d$ for that phonon. The equilibrium position of the atom is denoted by $\textbf{r}_{\boldsymbol{\ell}d}^0$. 
Using $\textbf{r}_{\boldsymbol{\ell}d} = \textbf{r}_{\boldsymbol{\ell}d}^0 + \textbf{u}(\boldsymbol{\ell}d)$ inside \eqref{eq:structurefactor}, the dynamic structure factor can be calculated in the harmonic approximation. This approach has been applied to calculate single-phonon excitations using numerical results for phonon energies and eigenvectors~\cite{Griffin:2018bjn,Trickle:2019nya,Cox:2019cod,Griffin:2019mvc,Trickle:2020oki,Griffin:2020lgd,Coskuner:2021qxo}, but becomes computationally much more burdensome for multi-phonons in the final state.

Under both the incoherent and harmonic approximations, it is possible to compute the multiphonon structure factor in \eqref{eq:structure_factor_inc}. This was given in Ref.~\cite{Campbell-Deem:2022fqm} as an expansion in the number of phonons produced $n$,
\begin{align}
\label{eq:structure_factor_crystal_harmonic}
S(\textbf{q},\omega) \approx & \ 2\pi \sum_{d} n_d \, |f_{d}|^2 e^{-W_d(\textbf{q})}   \sum_n \frac{1}{n!} \Big(\frac{q^2}{2m_d}\Big)^n \nonumber \\ & \times \left(\prod\displaylimits_{i=1}^n \int d \omega_{i} \frac{D_d(\omega_i)}{\omega_i} \right) \delta \left(\sum_{j=1}^n \omega_j - \omega \right),
\end{align}
where $D_d(\omega)$ is the partial density of states in the crystal, normalized to $\int d\omega D_d(\omega) = 1$. $W_d (\textbf{q})$ is the Debye-Waller factor defined as,
\begin{align}
W_d (\textbf{q}) = \frac{q^2}{4m_d}\int d\omega'~\frac{D_d (\omega')}{\omega'}.
\end{align}
\eqref{eq:structure_factor_crystal_harmonic} shows that with higher momentum $q$, there is an increased rate of multiphonons; the typical phonon number is $n \sim \frac{q^2}{2 m \bar \omega}$ with $\bar \omega$ a typical phonon energy. In the limit of $n \gg 1$, this reproduces the nuclear recoil limit.

In the incoherent approximation above, we still assumed the final states $|\Phi_f\rangle$ are the phonon eigenstates of the harmonic lattice Hamiltonian in \eqref{eq:harmonic_hamiltonian}. Let us now make a further approximation that  the final states are isolated atomic states, where each atom is bound in a potential. Assuming an isotropic potential, and a single frequency $\omega_0$ for the oscillators, a toy atomic Hamiltonian for atom $d$ in the lattice can be written as,
\begin{align}
H_{d}^{\text{toy}} = \frac{p_{d}^2}{2m_{d}} + \frac{1}{2} m_{d} \omega_0^2 r_{d}^2,
\end{align}
where ${\bf r}_{d}$ is the displacement of the atom $d$ from its equilibrium position. Following \eqref{eq:structure_factor_inc}, the dynamic structure factor can be written as,
\begin{align}
 S_{\text{toy}}(\textbf{q},\omega) &=  2\pi \sum_{d} n_d |f_{d}|^2 \sum_n \Big|\langle \vec{n}|e^{i\textbf{q} \cdot \textbf{r}_{d}}|0\rangle\Big|^2 \nonumber \\ & \times \delta(E_n - E_0 -\omega),
\end{align}
where $|\vec{n}\rangle$ are the energy eigenstates of the toy harmonic Hamiltonian considered for atom $d$, with $\vec{n}=\{n_x,n_y,n_z\}$. The energies with respect to the ground state equilibrium are given by $E_n - E_0 = n\omega_0$ with $n= n_x+n_y+n_z$. We have also absorbed the sum over the lattice vector $\boldsymbol{\ell}$ and the volume $V$ into the density $n_d$ of atom $d$ in the lattice. As shown in~\cite{Kahn:2020fef}, this structure factor is given by,
\begin{align}\label{eq:strcture_factor_toy_harmonic}
S_{\text{toy}}(\textbf{q},\omega) &\approx 2\pi\sum_{d} n_d|f_{d}|^2 e^{-2 W_d^{\text{toy}}(q)} \nonumber \\ &\times  \sum_n \frac{1}{n!} \left(\frac{q^2}{2 m_d \omega_0} \right)^n \delta \left( n \omega_0 - \omega \right),
\end{align}
where the Debye-Waller factor in the toy model is given by, $W_d^{\text{toy}}(q) = q^2/4m_d\omega_0$.

\begin{figure*}[t]
\centering
\includegraphics[width=0.98\linewidth]{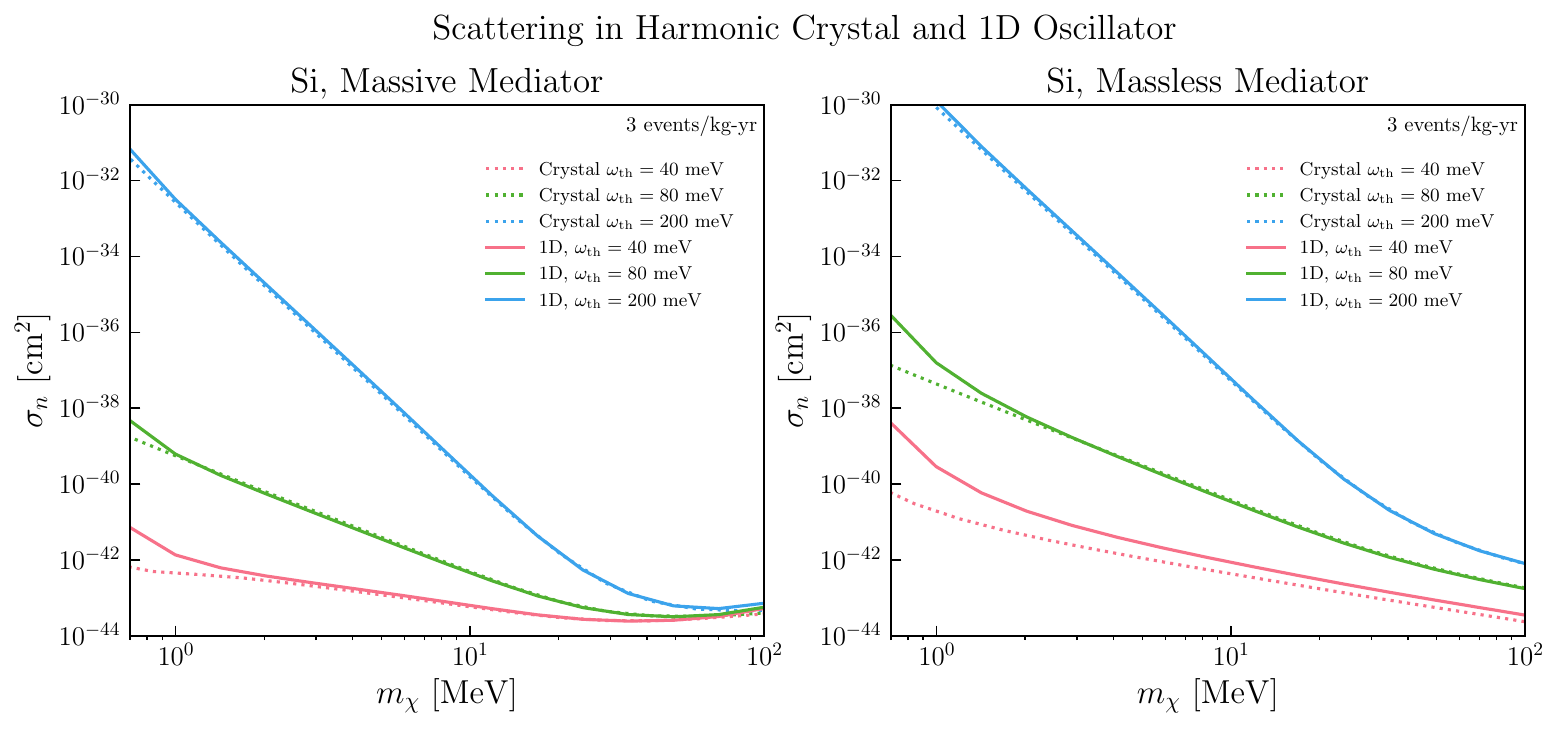} 
\caption{ 
\label{fig:toy_model_comparison} \textbf{Comparison of scattering in a harmonic crystal to 1D harmonic oscillator.} The dotted lines show the DM cross section reach computed using the multiphonon structure factor in a harmonic crystal, \eqref{eq:structure_factor_crystal_harmonic}, and assuming the incoherent approximation~\cite{Campbell-Deem:2022fqm}. Using the structure factor of the toy 1D harmonic oscillator in  \eqref{eq:strcture_factor_toy_harmonic} combined with the energy smearing prescription in \eqref{eq:replacement_prescription_approx} gives a very similar result (solid lines).
There are some small deviations at low momentum since we place a hard cut on the allowed momentum transfer $q > 2\pi/a \approx$ 2 keV for the 1D oscillator.}
\end{figure*}
This picture can be simplified even further by considering a toy one-dimensional harmonic potential for the atom $d$ given by 
\begin{align}
\label{eq:1Dho}
V_d(x) = \frac{1}{2}m_d\omega_0^2 x^2.
\end{align}
Note that in general $\omega_0$ will depend on the atom $d$ within the unit cell, but we suppress this dependence for simplicity. The structure factor in this 1D case is exactly the same expression as the toy three-dimensional case in \eqref{eq:strcture_factor_toy_harmonic}, as expected given the isotropic 3D potential assumed. A derivation of the 1D result is given in Sec.~\ref{sec:toy_harmonic_analytic}.

The toy model of DM scattering off a 1D harmonic potential gives a simple intuitive picture for the result in \eqref{eq:structure_factor_crystal_harmonic}.
We see a very similar form of the structure factor in \eqref{eq:strcture_factor_toy_harmonic}, but with a discrete spectrum of states for the isolated oscillator of the toy model. By assuming that the final states are isolated atomic states, we have effectively neglected the interactions between atoms, and the excited states of all the atoms are discrete and degenerate. In a real material, the interaction with neighboring atoms will lead to a splitting of the degenerate levels, and give a broad spectrum of allowed energy levels (the phonon spectrum). The interpretation for the structure factor is therefore also somewhat different in the two cases, as it gives a probability for exciting the $n$th excited state in an isolated oscillator. But we will still continue to refer the $n$th excited state as the $n$-phonon state to make the connection with the full incoherent structure factor in \eqref{eq:structure_factor_crystal_harmonic}.

The similarity in the structure factor gives a route forward to including anharmonic effects, which is much easier to understand in the toy model. We can proceed by including anharmonic corrections to the 1D potential in \eqref{eq:1Dho}, and in some cases obtain analytic results that illustrate their importance. In order to quantitatively estimate the impact on dark matter scattering rates, a few remaining ingredients are needed. In practice, the toy model can give very different results in certain parts of parameter space due to the discrete spectrum assumed and depending on the choice of $\omega_0$. We therefore need a prescription to identify the appropriate $\omega_0$ for the isolated oscillator, and to smear it out appropriately to mimic a real material.

Comparing Eqs.~\ref{eq:structure_factor_crystal_harmonic} and~\ref{eq:strcture_factor_toy_harmonic}, we see that the complete structure factor can be attained by making a replacement
\begin{equation}
    \label{eq:replacement_prescription}
    \frac{\delta(n \omega_0 - \omega) }{\omega_0^n}\rightarrow \left(\prod\displaylimits_{i=1}^n \int d \omega_{i} \frac{D(\omega_i)}{\omega_i} \right) \delta \left(\sum_{j=1}^n \omega_j - \omega \right).
\end{equation}
In this expression, we can identify $D(\omega)/(\omega \overline{\omega^{-1}})$ as a normalized probability distribution for $\omega$, where $\overline{\omega^{-1}} = \int d\omega' D(\omega')/\omega'$. This distribution yields a mean value for $\omega$ of $(\overline{\omega^{-1}})^{-1}$.
The right hand side of \eqref{eq:replacement_prescription} is proportional to the joint probability distribution for total energy $\omega$, and we can simplify it when $n \gg 1$ by applying the Central Limit Theorem. This allows us to replace the right hand side with a Gaussian, which simplifies  computations:
\begin{equation}
\label{eq:replacement_prescription_approx}
    \frac{\delta(n \omega_0 - \omega) }{\omega_0^n}\rightarrow \frac{ \big(\overline{\omega^{-1}}\big)^n  }{\sqrt{2\pi n \sigma^2}} e^{- \frac{\left(\omega - n \overline{\omega^{-1}}^{-1} \right)^2}{2 n \sigma^2}} \Theta(\omega_{\mathrm{max}} - \omega).
\end{equation}
Note we have included a cutoff at multiples of the maximum allowed energy in the density of states, $\omega_\mathrm{max} = n \times (\mathrm{min}(\omega) \vert D(\omega) = 0 )$ so that we do not include the region where $D(\omega_i) = 0$ on the right hand side of \eqref{eq:replacement_prescription}. The width of the Gaussian for $n=1$ is given by
\begin{equation}
    \sigma = \sqrt{\frac{\overline{\omega}}{\overline{\omega^{-1}}} - \frac{1}{\big(\overline{\omega^{-1}} \big)^2}}
\end{equation}
and $\overline{\omega} = \int d\omega' D(\omega') \omega'$. This discussion therefore makes it clear that we should identify the frequency of the 1D toy model as $\omega_0 = 1/\overline{\omega^{-1}}$, which can be calculated numerically given the phonon density of states. This approach is validated in Fig.~\ref{fig:toy_model_comparison}, where we compare our previous result using the full density of states \cite{Campbell-Deem:2022fqm} to the prescription described above. Note that small deviations at low mass arise from the lack of a cutoff at the Brillouin zone momentum in the previous density of states result. We reiterate that in this work, we shall include this Brillouin zone cutoff across all rate calculations since the incoherent approximation and subsequent approximations are only valid in this regime.

We will utilize this prescription to extend the multiphonon calculations for an anharmonic potential.  To set up toy 1D anharmonic potentials, we first need to understand the anharmonic properties of typical crystals to extract the behavior of the potentials. We do this in the following subsection.

\subsection{Anharmonic crystal properties}
\label{sec:anharmonic_properties}

In general, a crystal lattice will  exhibit some anharmonicity. Anharmonicity technically refers to the presence of non-zero force constants which are higher than second-order in the lattice potential in \eqref{eq:latticepotential}. For example, cubic anharmonicity in the crystal is parameterized by the third-order force constants $k^{(3)}_{\alpha\beta\gamma}(\boldsymbol{\ell}d,\boldsymbol{\ell}'d',\boldsymbol{\ell}''d'')$ in \eqref{eq:latticepotential}. Such force constants can be computed with DFT methods, similar to the harmonic case~\cite{DEBERNARDI1994813}. In the presence of such terms, the phonon eigenstates are no longer the harmonic phonon eigenstates of the crystal, and higher order phonon interactions, such as a three-phonon interaction, will be present. Calculating the full dynamic structure factor in \eqref{eq:structure_factor_inc} for a crystal with such anharmonicity would require accounting for these higher order force tensors in both the matrix elements and in the final states, which quickly becomes a very challenging numerical problem. The rough size of the anharmonic force constants can be inferred from measurable crystal properties, however. We will briefly discuss some of the anharmonic effects below, and use them to justify our estimate of anharmonic effects. 

An important effect of keeping cubic or higher order terms in \eqref{eq:latticepotential} is to introduce interactions between the phonon modes which are the eigenstates of the harmonic Hamiltonian. For example, from \eqref{eq:displacementoperators}, we can see that a cubic term in the displacements $\textbf{u}(\boldsymbol{\ell}d)$ will introduce three-phonon interactions like $\hat{a}_{\textbf{q},\nu}^{\dagger}\hat{a}_{\textbf{q}',\nu'}\hat{a}_{\textbf{q}'',\nu''}$ (i.e. annihilation of two phonons to create a single phonon) or $\hat{a}_{\textbf{q},\nu}^{\dagger}\hat{a}_{\textbf{q}',\nu'}^{\dagger}\hat{a}_{\textbf{q}'',\nu''}$ (i.e. decay of a single phonon into two phonons)  in the Hamiltonian at the first order in the anharmonic force constant $k^{(3)}$. Phonon lifetimes in crystals are thus directly related to the anharmonic force constants, and can be measured to estimate the size of the anharmonicity~\cite{PhysRevB.102.174311,Wei_2021,PhysRevLett.110.157401}. 

Anharmonicity is also necessary to explain thermal expansion and conductivity in crystals. In particular, the linear volume expansion coefficient of crystals can be directly written in terms of the mode Gruneisen constants $\gamma_{\textbf{q}\nu}$ which is defined for phonon modes labelled by the momentum $\textbf{q}$ and branch index $\nu$ as~\cite{mayer1984calculation},
\begin{align}
\gamma_{\textbf{q}\nu} = -\frac{V}{\omega_{\textbf{q}\nu}}\frac{\partial \omega_{\textbf{q}\nu}}{\partial V}.
\end{align}
Note that the change in volume in the equation above is at a fixed temperature. In a purely harmonic crystal, the phonon frequencies are determined by the second-order force constants which do not get modified with changes in volume, thus leading to zero Gruneisen constant. However, in the presence of cubic anharmonicity, the phonon frequencies are determined by the effective second-order force constants, which receive corrections depending on both the third-order force constants $k^{(3)}$ and the changes in volume, thus giving a non-zero Gruneisen constant~\cite{Lee_2017}. An increase in volume leads to larger displacements of atoms, which typically makes the effective second order constants and the phonon frequencies smaller, providing a positive Gruneisen constant. In the case of a non-zero Gruneisen constant, the free energy of the crystal, which has a harmonic contribution $\propto \Delta V^2$, receives a volume-dependent correction $\propto -\Delta V \gamma_{\textbf{q}\nu} \bar{E}_{\textbf{q}\nu}$, where $\bar{E}_{\textbf{q}\nu}$ is the mean energy in the phonon mode $\textbf{q}\nu$ at a particular temperature~\cite{srivastava2019physics}. As the temperature increases, the mean energy $\bar{E}_{\textbf{q}\nu}$ goes up, and thus this leads to a new equilibrium volume which minimizes the free energy. For a positive Gruneisen constant, this leads to thermal volume expansion. 

The Gruneisen constants are thus directly related to the cubic force constants of the material, and have also been used to extract them~\cite{Lee_2017}. Concretely, the relationship between the mode Gruneisen constants and the anharmonic force constants for weak anharmonicity can be shown to be~\cite{PhysRevB.84.085204},
\begin{align}\label{eq:Gruneisen}
\gamma_{\textbf{q}\nu} &= -\frac{1}{6\omega_{\textbf{q},\nu}^2}\sum_{d,\boldsymbol{\ell}'d',\boldsymbol{\ell}''d''}\sum_{\alpha \beta \delta}  k_{\alpha\beta\delta}^{(3)}(\textbf{0}d,\boldsymbol{\ell}'d',\boldsymbol{\ell}''d'')\nonumber \\& \times \frac{e_{\textbf{q},\nu}^{\beta} (d')^{*}e_{\textbf{q},\nu}^{\delta} (d'')}{\sqrt{m_{d'}m_{d''}}}~r_{\textbf{0}d}^{0,\alpha}~e^{i\textbf{q} \cdot (\boldsymbol{\ell}''-\boldsymbol{\ell}')},
\end{align}
where the $e_{\textbf{q},\nu}^{\beta}(d)$ indicates the displacement of atom $d$ in the Cartesian direction $\beta$ for the phonon $\textbf{q}\nu$, and $r_{\textbf{0}d}^{0,\alpha}$ is the equilibrium position of atom $d$ in the Cartesian direction $\alpha$ for the unit cell at the origin. To get a rough estimate of the maximum anharmonicity strength in the crystal, the relation in \eqref{eq:Gruneisen} can be inverted and written in terms of the maximal mode Gruneisen constant $\gamma^{\text{max}}$ found in a crystal,
\begin{align}\label{eq:cubic_gruneisen}
k^{(3)} \sim \frac{6 m_d \omega_0^2 \gamma^{\text{max}}}{l},
\end{align}
where $\omega_0$ is the typical phonon energy of the lattice and $l$ is the nearest neighbor distance. Now consider a typical displacements $\sim (\sqrt{2m_d\omega_0})^{-1}$ of an atom in the crystal; the change in the potential energy $\delta V_{\text{anh}}$ due to anharmonic force constant estimated above is given by,
\begin{align}\label{eq:general_cubic_estimate}
\frac{\delta V_{\text{anh}}}{\omega_0} &\sim \frac{1}{\omega_0}\frac{1}{3!}~k^{(3)} (\sqrt{2m_d\omega_0})^{-3} \nonumber\\& \sim 0.02 ~ \Big(\frac{m_d}{28~\text{GeV}}\Big)^{-0.5}~\Big(\frac{\omega_0}{30~\text{meV}}\Big)^{-0.5}\nonumber\\& \times\Big(\frac{\gamma^{\text{max}}}{1.5}\Big)~\Big(\frac{l}{2.35~\AA}\Big)^{-1},
\end{align}
where in the second line we use parameters for Si. We use an estimate for the maximal value of the mode Gruneisen constant in Si from~\cite{srivastava2019physics} at 0K. In Ge, the maximal Gruneisen constant is similar to that in Si, while in GaAs, it could be as high as 3.5 for certain phonon modes~\cite{srivastava2019physics}. The Gruneisen constant thus provides a rough estimate of the overall anharmonicity in the crystal, including the cubic terms which depend on displacements of multiple atoms.

In this paper, we will work with a toy model of anharmonic interactions similar to the 1D oscillator model in Sec.~\ref{sec:structure_factor_harmonicToToy}. In particular, we consider excitations for an isolated atom in a 1D anharmonic potential. The anharmonicity is controlled by force constant terms like $k^{(3)}_{\alpha\beta\gamma}(\boldsymbol{\ell}d,\boldsymbol{\ell}'d',\boldsymbol{\ell}''d'')$ with $\boldsymbol{\ell}d=\boldsymbol{\ell}'d'=\boldsymbol{\ell}''d''$ which characterize the modification to the potential of a single atom in a lattice. Since the Gruneisen constants involve a sum over many cubic force terms, we instead directly obtain the single-atom anharmonic force constants with an empirical model of the lattice.

\begin{figure}[t]
	\centering
\includegraphics[width=\hsize]{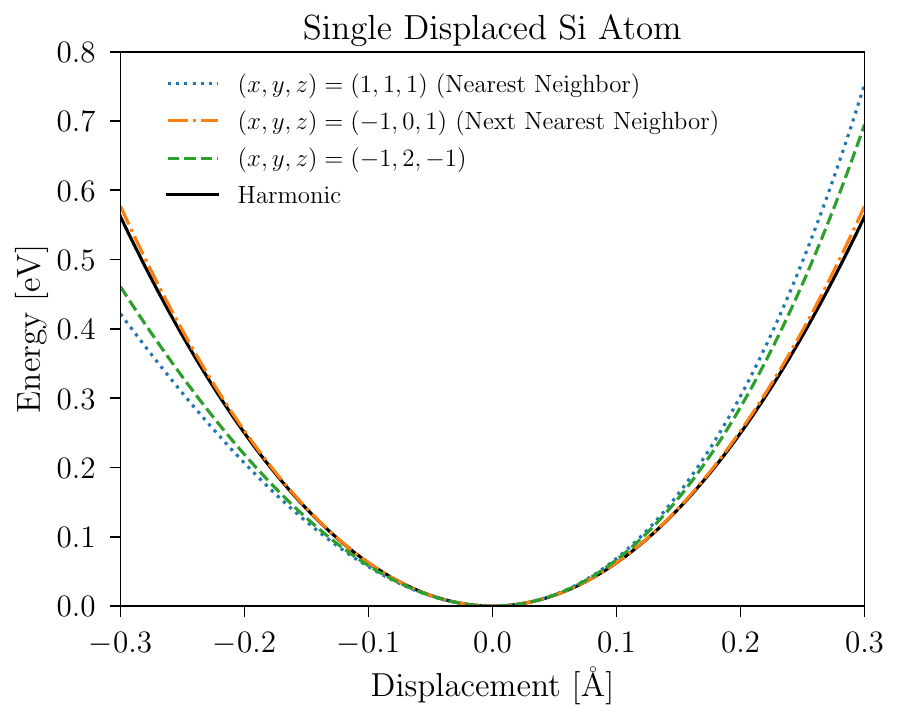}
	\caption{\textbf{Single atomic potential:} Potential of a single atom displaced along various directions with all other atoms at their equilibrium positions. In zincblende Si, the largest anharmonicity is in the direction of the nearest-neighbor atom, while the smallest anharmonicity is in the direction of the next-nearest-neighbor. We have also included a third direction orthogonal to the other two, with intermediate anharmonicity strength. }
	\label{fig:atomic_potential}
\end{figure}

We model the lattice assuming empirical interatomic potentials, which have been shown to accurately reproduce phonon dispersions and transport properties~\cite{Rohskopf:2017}. Concretely, we assume the Tersoff-Buckingham-Coulomb interatomic potential with the parameter set given in Ref.~\cite{Rohskopf:2017} (see Appendix~\ref{appendix:interatomic_potentials} for details). We then fix all atoms at their equilibrium positions except for one atom denoted by $\boldsymbol{\ell}d$, which is displaced by a small distance in different directions. The single atom potential calculated from this procedure is shown in Fig.~\ref{fig:atomic_potential} for Si, with deviations from the harmonic potential that depend on the direction of displacement. The maximum anharmonicity is along the direction of the nearest neighbor atom. Along this direction, we find that the typical change in the potential energy for an atom displaced by $r \sim (\sqrt{2m_d\omega_0})^{-1}$ is,
\begin{align}
\frac{\delta V_{\text{anh}}}{\omega_0} \sim 0.01. 
\end{align}
Comparing this estimate with \eqref{eq:general_cubic_estimate}, we see that the anharmonicity strength inferred from the potential of a single atom is roughly of the same size as the overall anharmonicity strength of the lattice inferred from the Gruneisen constant. Thus, even though we do not perform a full calculation of the structure factor for an anharmonic crystal including the modification of the phonon spectrum and the lattice states, the comparison above suggests that the effects in a full calculation are expected to be similar in magnitude to the effects we estimate in this work using single atom potentials.

\subsection{Toy anharmonic potential}
\label{sec:toy_anharmonic}

As shown in Sec.~\ref{sec:structure_factor_harmonicToToy} for the harmonic crystal, the features of the dynamic structure factor under the incoherent approximation can be well-approximated with just a 1D toy potential for an individual atom. This gives a much simpler path to calculating DM scattering in anharmonic crystals for $q \gg 2\pi/a$, where many phonons may be produced. In contrast, prior work including anharmonicity focused on the limit $q \ll 2\pi/a$, restricted to two phonons~\cite{Campbell-Deem:2019hdx}, and does not scale well to large number of phonons. We can then stitch together the two approaches to gain a more complete understanding of anharmonic effects.  

In this work, we take a 1D anharmonic potential and calculate the 1D structure factor, in order to simplify the problem as much as possible.  Taking the 1D approximation is more subtle in the presence of anharmonicity since a generic potential in 3D is not separable, unlike the harmonic case. Denoting the small displacement around equilibrium by $r$, and the polar and azimuthal directions by $\theta$ and $\phi$ respectively, the potential energy for atom $d$ in the lattice can be expanded in powers of $r$ as,
\begin{align}
    \label{eq:single_atom_potential}
    V_d(r, \theta, \phi)  = &\ \ \frac{1}{2} m_d \omega_0^2 r^2 \nonumber \\
    & + \sum_{k \geq 3} \lambda_k \omega_0 f_k(\theta, \phi) (r \sqrt{2 m_d \omega_0} )^k, 
\end{align}
where $\lambda_k$ are dimensionless constants parameterizing the degree of anharmonicity at $k^{\text{th}}$ order, and  $f_k(\theta,\phi)$ are functions which specify the angular dependence and whose range is $[-1, 1]$. Solving the full 3D problem would require numerically finding the eigenstates of this general potential, while in the 1D case we can make much more progress analytically. We will therefore select directions of maximum anharmonicity and use this for our simplified 1D problem. Our expectation is that this gives a conservative estimate of the importance of anharmonic couplings, in that the full 3D calculation would give somewhat reduced effects.  

As discussed in Sec.~\ref{sec:anharmonic_properties}, we can extract realistic single atom potentials by modeling the interatomic potentials on the lattice and displacing a single atom (see Appendix~\ref{appendix:interatomic_potentials} for details). We typically find that, for small displacements around equilibrium, the anharmonicity is dominated by the cubic and quartic terms parametrized by $\lambda_3$ and $\lambda_4$, respectively. Motivated by these observations, we consider the following forms of toy potentials in our study:
\begin{itemize}
\item \textbf{Single cubic or quartic perturbations:} We first consider a harmonic potential with a single perturbation,
\begin{align}
\label{eq:toy_anharmonic_potential1}
V_d(x) = \frac{1}{2} m_d \omega_0^2 x^2 + \lambda_k\omega_0 (\sqrt{2 m_d \omega_0} x)^k,
\end{align}
where $k =3$ or 4. This case is amenable to perturbation theory, and in Sec.~\ref{sec:perturbation_theory}, we apply it to discuss the power counting of anharmonic corrections.
\item \textbf{Morse potential:} It is possible to obtain exact (non-perturbative) analytic results for the Morse potential defined by,
\begin{equation}
    V_{\mathrm{Morse}} (x) =  B \Big( e^{-2 a x} - 2 e^{-a x} \Big),
\end{equation}
where $a$ is a parameter controlling the width of the potential and $B$ is the normalization. We fit these two parameters to the cubic anharmonicity estimated from the single atom potentials discussed earlier, and calculate the dynamic structure factor for this potential in App.~\ref{sec:morse}. 
\item \textbf{Fit to realistic atomic potentials:} We numerically calculate the structure factor in a potential with both cubic and quartic terms, where the dimensionless anharmonic couplings are obtained by fitting to the actual single atom potential. The potential in this case is given by
\begin{align}
V_d(x) &= \frac{1}{2} m_d \omega_0^2 x^2 + \lambda_3\omega_0 (\sqrt{2 m_d \omega_0} x)^3 \nn 
\\
&\quad + \lambda_4\omega_0 (\sqrt{2 m_d \omega_0}x)^4.\label{eq:toy_anharmonic_potential}
\end{align}
We find that typically, $\lambda_3 \sim 0.01$, and $\lambda_4 \sim 10^{-4}$.
\end{itemize}

For the 1D toy potentials discussed above, we compute the 1D dynamic structure factor in the incoherent approximation ($ q\gg 2\pi/a$):
\begin{align}
\label{eq:structure_factor_toy_1D}
S_{\text{toy}}(q,\omega) &= 2\pi \sum_{d} n_d ~|f_d|^2 \sum_f \left|    \mel{\Phi_f}{e^{iqx}}{\Phi_0} \right|^2 \nonumber \\& \times \delta (E_f - E_0 - \omega).
\end{align}
Again, we have summed over all atoms of type $d$ in the lattice and defined the number density of atom $d$ by $n_d$. 
The wavefunctions $|\Phi\rangle$ are the eigenfunctions of the Hamiltonian,
\begin{align}
H_{\text{toy}} = \frac{p^2}{2m_d} + V_d (x).
\end{align}
The computation of the dynamic structure factor then boils down to computing the ground state $|0\rangle$ and the excited eigenstates $|\Phi_f \rangle$ for this Hamiltonian, and calculating the structure factor under the incoherent approximation as in Eq.~\eqref{eq:structure_factor_toy_1D}.

As discussed in Sec.~\ref{sec:structure_factor_harmonicToToy}, for a 1D toy model the phonon levels are discrete and in a real crystal there is a broad spectrum of energy levels. Similar to the harmonic case, we need a prescription to account for this smearing of energies. In the case with anharmonicity, the spectrum is shifted. The 1D toy model will instead give a modified energy-conserving delta function:
\begin{equation}
    \delta( f(n)\omega_0 - \omega),
\end{equation}
where $f(n) \omega_0$ is the energy difference between the $n$th excited state and the ground state. $f(n)$ will depend on the exact form of the potential. Guided by the harmonic result, we again shall fix $\omega_0 = 1/\overline{\omega^{-1}}$ and introduce a width to the delta function in a similar fashion:
\begin{equation}
\label{eq:replacement_prescription_anharmonic}
    \delta(f(n)\omega_0 - \omega) \rightarrow  \frac{1}{\sqrt{2\pi f(n) \sigma^2}} e^{- \frac{(\omega - f(n) \omega_0)^2}{2 f(n) \sigma^2}}.
\end{equation}
This is in the 1D approximation, and that including the full 3D anharmonic potential would be expected to have an additional effect on the spectrum of states. However, in practice the anharmonicity is sufficiently small that the shift of the spectrum is subdominant to the other anharmonic effects in the structure factor.

This forms the basis of the toy model we consider in this paper. Focusing on the high $q$ regime where the incoherent approximation applies, we consider independent lattice sites and calculate scattering in them with 1D toy anharmonic potentials. We now describe different approaches to understand the dynamic structure factor in this setting.

\section{Analytic results for structure factor}\label{sec:analytic}

In this section, we study the features of the structure factor for a 1D anharmonic potential with analytic methods. This will allow us illustrate the general behavior for the limits $q \ll \sqrt{2 m_d \omega_0}$ and $q \gg \sqrt{2 m_d \omega_0}$. 

First, we review the derivation of the structure factor for a 1D harmonic potential. For $n$-phonon production in the harmonic limit, the structure factor in the regime $q \gg \sqrt{2m_d\omega_0}$ is $\propto q^{2n}/(2m_d \omega_0)^n$. Treating the anharmonic 1D potential as a perturbation, we then show that the $q$ dependence of the $n$-phonon term can be substantially modified in the regime $q \ll \sqrt{2m_d\omega_0}$, leading to large anharmonic corrections. In particular, we obtain the power counting of the structure factor in powers of $q$ and the anharmonicity parameter $\lambda_k$, which allows us to roughly identify the regime of $q$ where we expect the anharmonic effects to be dominant. As we will see later, this proves useful to explain the numerical results for realistic potentials. 

Finally, we will also use the impulse approximation to perform an analytic estimate of the structure factor in the regime $q > \sqrt{2m_d\omega_0}$. We show that the nuclear recoil limit is reproduced, with the structure factor approximated by a Gaussian envelope similar to the harmonic case. Anharmonic terms give rise to slightly modified shape of the Gaussian, which have negligible impact on scattering rates. 

\subsection{Harmonic oscillator}\label{sec:toy_harmonic_analytic}

First, we briefly review the calculation of the dynamic structure factor in the harmonic approximation. In this case the potential $V_d (x)$ is given by
\begin{align}
V_d (x) = \frac{1}{2}m_d \omega_0^2 x^2.
\end{align}
The energy $E_n$ of the $n$-th excited state $|n\rangle$ of this simple harmonic oscillator is given by,
\begin{align}
E_n = \Big(n+\frac{1}{2}\Big)\omega_0.
\end{align}
The structure factor in Eq.~\eqref{eq:structure_factor_toy_1D} thus becomes,
\begin{align}
\label{eq:structure_factor_toy_1D_Har}
S_{\text{toy}}(q,\omega) = 2\pi \sum_{d} n_d ~ |f_{d}|^2 \sum_n \left|    \mel{n}{e^{iqx}}{0} \right|^2 \delta (n \omega_0 - \omega).
\end{align}
The matrix element can be evaluated in the following way,
\begin{align}\label{eq:0nmatrixelement}
\mel{n}{e^{iqx}}{0} &= \frac{1}{\sqrt{n!}}\mel{0}{a^n e^{iqx}}{0} \nonumber \\ &= \frac{1}{\sqrt{n!}}\mel{0}{e^{iqx} \Big(a+\frac{iq}{\sqrt{2m_d\omega_0}}\Big)^n}{0} \nonumber \\ &= \frac{1}{\sqrt{n!}}\Big(\frac{iq}{\sqrt{2m_d\omega_0}}\Big)^n \mel{0}{e^{iqx}}{0} \nonumber \\&= \frac{1}{\sqrt{n!}}\Big(\frac{iq}{\sqrt{2m_d\omega_0}}\Big)^n e^{-\frac{q^2}{4m_d\omega_0}},
\end{align}
where we use $e^{-iqx} a e^{iqx} = a+\frac{iq}{\sqrt{2m_d \omega_0}}$ in the second equality.
Plugging the above matrix element to the structure factor in \eqref{eq:structure_factor_toy_1D_Har} becomes,
\begin{align}
S_{\text{toy}}(q,\omega) &= 2\pi \sum_{d}n_d ~ |f_{d}|^2 e^{-2 W_d^{\text{toy}}(q)} \nonumber \\ &\times  \sum_n \frac{1}{n!} \left(\frac{q^2}{2 m_d \omega_0} \right)^n \delta \left( n \omega_0 - \omega \right),
\label{eq:1dsqw_harmonic}
\end{align}
where $W_d^{\text{toy}}(q)= q^2/(4m_d\omega_0)$ is the Debye-Waller factor in the toy model. The structure factor follows a Poisson distribution with mean number of phonons $\mu = q^2/(2 m_d \omega_0)$, as also shown in the case of the 3-dimensional harmonic oscillator in ~\cite{Kahn:2020fef}. 
\\

\subsection{Perturbation theory for anharmonic oscillator: $q \ll \sqrt{2 m \omega_0}$}\label{sec:perturbation_theory}

We now turn to more general case where small anharmonic terms are included in the 1D toy potential. An exact solution is no longer possible. But as we will see, in the kinematic regime $q \ll \sqrt{2 m_d \omega_0}$, we can use perturbation theory to obtain the behavior of the structure factor and illustrate the importance of the anharmonic corrections as a function of momentum and energy deposition. Our goal in this section then is to obtain the power counting of the anharmonic contributions to the structure factor.

The toy Hamiltonian we consider is given by,
\begin{align}\label{eq:toy_hamiltonian_perturbation}
H_{\text{toy}} = \frac{p^2}{2m_d} + \frac{1}{2}m_d \omega_0^2 x^2 + \lambda_k\omega_0 (\sqrt{2 m_d \omega_0} x)^k.
\end{align}
We will concretely consider $k$ equal to 3 and 4, corresponding to a leading cubic and quartic anharmonicity, respectively.  
Treating the dimensionless anharmonicity parameter $\lambda_k$ as a perturbation,  the eigenstates $|\Phi_n\rangle$ are given by
\begin{align}\label{eq:perturbedeigenstate}
|\Phi_n\rangle = |n\rangle + \lambda_k~|\psi_n^{(1)}\rangle+ \lambda_k^2~|\psi_n^{(2)}\rangle +... ,
\end{align}
and $E_n'$ are the perturbed energies,
\begin{align}\label{eq:energyshiftdef}
E_n' = \Big(n+\frac{1}{2}\Big)\omega_0 + \lambda_k~ c_n^{(1)} + \lambda_k^2~ c_n^{(2)} +...
\end{align}

With time-independent perturbation theory, the dynamic structure factor can be explicitly computed at different orders in $\lambda_k$ using \eqref{eq:structure_factor_toy_1D}. We defer the details of the explicit calculation to Appendix~\ref{appendix:perturbation_theory}. Instead, from the structure of the expansion we can already learn about the relevant corrections. In general, we can express the dynamic structure factor as an expansion in both $\lambda_k$ and $q^2/(2m_d\omega_0)$. At zeroth order in $\lambda_k$, we see from \eqref{eq:1dsqw_harmonic} that the $n$-phonon term appears with a $q$-scaling of $q^{2n}/(2m_d\omega_0)^n$. As we will show below, anharmonicity introduces departures from this $q$-scaling at higher orders of $\lambda_k$. In the kinematic regime under consideration ($q \ll \sqrt{2 m_d \omega_0}$), powers of $q^2/(2m_d\omega_0)$ smaller than $n$ can lead to large anharmonic corrections to the $n$-phonon term in the structure factor.\footnote{Perturbation theory in $\lambda_k$ is still valid. For instance, the expansion in \eqref{eq:perturbedeigenstate} still holds. But the harmonic contribution in the structure function could be suppressed by small $q$ for multi-phonon states.} The aim of this section is thus to illustrate the behavior of the $q$-scaling at different orders of $\lambda_k$.

The general expression for the dynamic structure factor in the toy model can be written as,
\begin{align}\label{eq:perturbation_expansion}
S_{\text{toy}} &(q,\omega) =  2\pi  \sum_{d} n_d ~|f_{d}|^2 e^{-2W^{\text{toy}}_d (q)}\times  \\ 
 & \sum_n  \delta(E_n' - E_0' - \omega) \Bigg[\frac{1}{n!}\Big(\frac{q^2}{2m_d\omega_0}\Big)^n \nonumber\\&+ \sum_{i \geq 1} \Big(\frac{q^{2}}{2m_d\omega_0}\Big)^{i}  \left( a_{n,i}~\lambda_k^{\nu(n,i)}~ + \mathcal{O}\Big(\lambda_k^{\nu(n,i)+1}\Big)\right)\Bigg] \nonumber
\end{align}
For each $n$, the harmonic contribution appears at $\mathcal{O}((q^{2}/(2m_d \omega_0))^n)$ as seen in \eqref{eq:strcture_factor_toy_harmonic}; note that we do not include the Debye-Waller factor in this power counting discussion since it always appears as an overall factor. The anharmonic corrections are included here as an expansion in powers of $q^2/(2m_d\omega_0)$ which are denoted by $i$. From the orthogonality of the states $|\Phi_n\rangle$ with the ground states, we see that the dynamic structure factor should vanish for $q \rightarrow 0$, which in turn implies that $i \geq 1$. Each power $i$ of $q^2/(2m_d\omega_0)$ appears with non-zero powers of $\lambda_k$, denoted by $\nu(n,i)$. Here the power $\nu(n,i)$ is the smallest \textit{allowed} power of $\lambda_k$ for a given phonon number $n$ and the power $i$ of $q^2/(2m_d\omega_0)$. However, numerical cancellations can sometimes force this leading behavior to vanish. Typically, the bigger the difference in $i$ and $n$, the larger  the power of $\lambda_k$ that is required. We will explicitly see the behavior of the powers $\nu(n,i)$ for $k$ equal to 3 and 4 below, but we first discuss the implications of this form.

For the single phonon structure factor (i.e. for $n=1$), the anharmonic terms are always suppressed compared to the harmonic term because of the additional powers of $\lambda_k$ and $q^2/(2m_d\omega_0)$. But for phonon numbers $n>1$, it is possible for anharmonic contributions to dominate for $q \ll \sqrt{2m_d\omega_0}$. 
As a simple example, in the 3-phonon state, the harmonic contribution to the structure factor is proportional to $q^{6}/(2m_d\omega_0)^3$, while the aharmonic result contains $\lambda_3^2 q^{4}/(2m_d\omega_0)^2$. So when $q \ll \sqrt{2m_d\omega_0}$, the anharmonic effect can lead to a large correction to the dynamic structure factor.

In a generic $n$-phonon state, the harmonic piece scales as $(q^2/(2m_d\omega_0))^n$.
Comparing this with the anharmonic term $\propto \lambda_k^{\nu(n,i)}q^{2i}/(2m_d\omega_0)^i$, we note that the anharmonic term dominates the harmonic term for $q \ll \sqrt{2m_d\omega_0}\lambda_k^{\nu(n,i)/(2(n-i))}$. For small enough $q$, the behavior is governed by the anharmonic effects. Of course, at even smaller $q \sim q_{\rm BZ}$ one would expect the incoherent approximation to break down. For the values of $\lambda$ in realistic materials, we find that the dominance of the anharmonic terms can happen for $q$ above $q_{\rm BZ}$, particularly for larger $n$. These corrections become larger with $n$ since the harmonic piece is progressively more suppressed in $q^2/(2m_d\omega_0)$.

We now illustrate the origin of the $\lambda_k$ powers $\nu(n,i)$ with an example in the case of $k=3$. In this case, the perturbation $x^3 \sim (a + a^\dagger)^3$ implies the leading correction to the state can change the oscillator number by $\pm 1$ or $\pm 3$. Then the perturbed eigenstates have the schematic form:
\begin{align}
|\Phi_n\rangle \sim |n\rangle &+ \lambda_3 \,\left(|n-3\rangle + |n-1\rangle + |n+1\rangle + |n+3\rangle \right) \nonumber\\
& +\mathcal{O}(\lambda_3^2).
\end{align}
We neglect the numerical prefactor in front of each state.
Note that the terms are only present if the integer labelling the state is non-negative, for example for the ground state $|\Phi_0\rangle \sim |0\rangle +~ \lambda_3~( |1\rangle + |3\rangle) +\mathcal{O}(\lambda_3^2)$.
The matrix element appearing in the $n$-phonon structure factor can be expressed as,
\begin{align}
\langle\Phi_n|e^{iqx}|\Phi_0\rangle &\sim b_0 + \lambda_3 b_1 + \lambda_3^2 b_2 +  \mathcal{O}(\lambda_3^3), 
\end{align}
where the coefficients are schematically given by,
\begin{align}
b_0 &\sim \langle n|e^{iqx}|0\rangle \\
b_1 &\sim \langle n-3|e^{iqx}|0\rangle~+ \nonumber\\& \langle n-1|e^{iqx}|0\rangle  + \langle n+1|e^{iqx}|0\rangle~+ \nonumber\\& \langle n+3|e^{iqx}|0\rangle + \langle n|e^{iqx}|1\rangle + \langle n|e^{iqx}|3\rangle
\end{align}
In order for given term in the coefficient to be nonzero, a minimum number of powers of $iqx$ are required in the series expansion for $e^{i qx}$. This therefore links the powers of $q$ with powers of $\lambda_3$. 

Taking $n = 3$ as an example, then $b_0 \propto (i q)^3 $  at leading order in the $q$ expansion. Meanwhile, $b_1 \propto (i q)^2 + (i q)^4 + ...$. Note that the matrix elements $\langle 0 | e^{iqx}|0\rangle$ and $\langle 3 | e^{iqx}|3\rangle $ in $b_1$ contain terms proportional to $(iq)^0$, but they cancel each other, consistent with a matrix element that always vanishes as $ q \to 0$. Also note that the coefficients $b_0, b_1$ always alternate in even or odd powers of $(i q x)$ and therefore alternate in being purely real or imaginary. The resulting matrix element squared thus goes as
\begin{align}
|\langle\Phi_3|e^{iqx}|\Phi_0\rangle|^2 &\sim |b_0 + \lambda_3^2 b_2 + \mathcal{O}(\lambda_3^4)|^2 + |\lambda_3 b_1 + \mathcal{O}(\lambda_3^3)|^2, \nonumber \\
& \sim q^
{6} + \lambda_3^2 (q^4 + \mathcal{O}(q^6)) + \mathcal{O}(\lambda_3^4).
\end{align}
For the cubic interaction, only even powers of $\lambda_3$ appear in the matrix element squared due to the alternating even and odd powers of $(i qx)$ in the $b$ coefficients. In this example, in order to achieve the minimum $q$ scaling of $q^2$, higher powers of $\lambda_3$ are required, which will introduce more terms in the expansion. Here we see a correction to the matrix element squared at $O(q^2 \lambda_3^4)$.

\begin{figure*}
    \centering \includegraphics[width=0.94\linewidth]{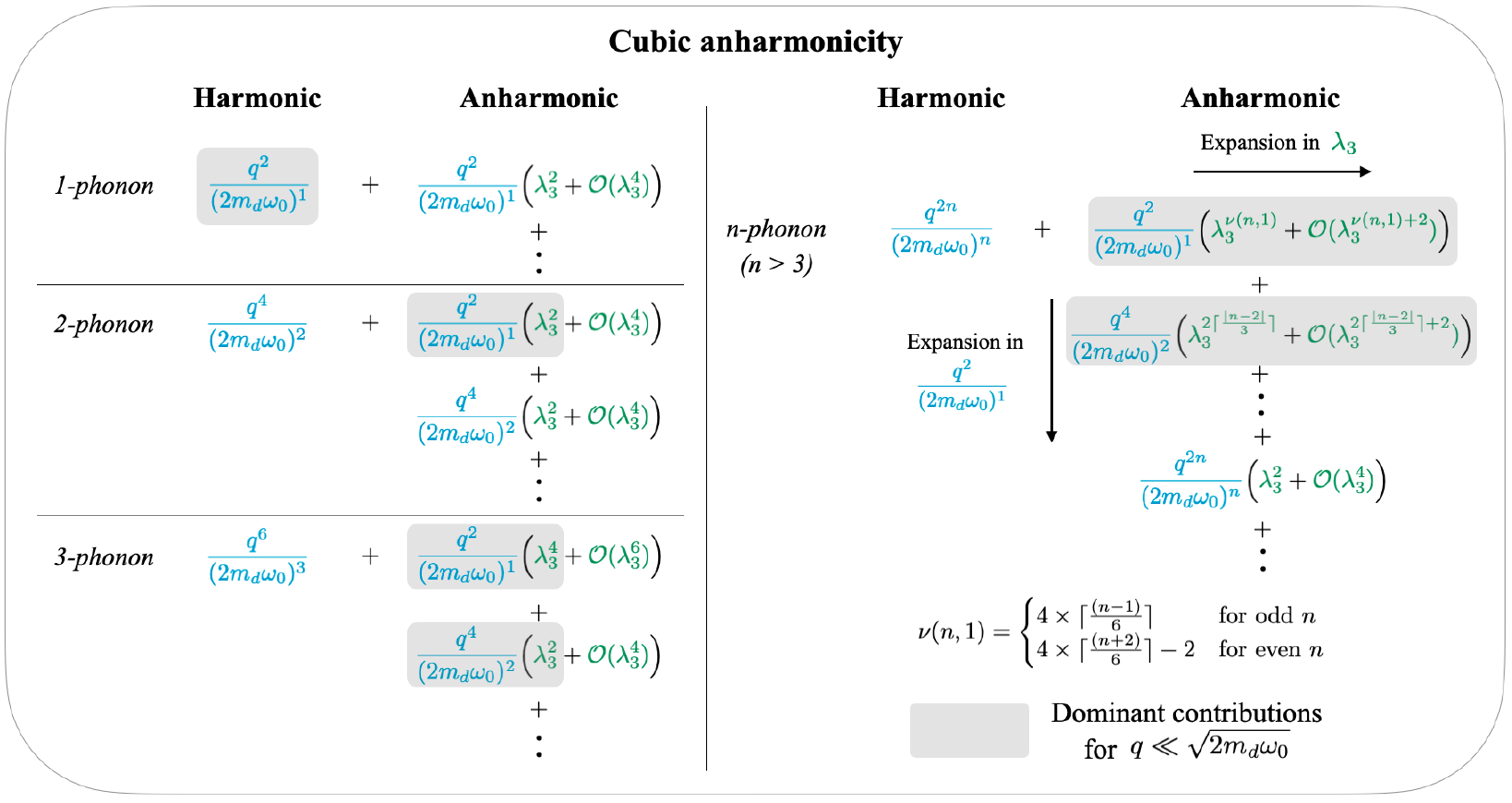}\caption{Expansion of the structure factor in phonon number $n$, powers of $q^2/(2m_d\omega_0)$, and powers of $\lambda_3$ for a cubic perturbation ($k=3$ in \eqref{eq:toy_hamiltonian_perturbation}). The right part shows the general behavior of the $n$-phonon term for $n>3$, while the left part shows the expansion for $n=$1, 2, and 3. Shaded terms show the dominant contributions when $ q\ll \sqrt{2 m_d \omega_0}$, which comes from the anharmonic terms for $ n \ge 2$. Here we just illustrate the power counting; individual terms might not be present if there is a numerical cancellation in the coefficients.  }
\label{fig:anharmonic_schematic} 
\end{figure*}

The explicit derivation of $\nu(n,i)$ is given in Appendix~\ref{appendix:perturbation_theory}. The minimum power of $\lambda_3$ required to get the leading behavior $\propto q^2/(2m_d\omega_0)$ in the anharmonic terms is given by,
\begin{equation} 
\nu(n,1) = 
\begin{cases}
    \text{max}\Big(4 \times \left\lceil \frac{(n-1)}{6} \right\rceil~,~2 \Big) & \text{for odd} ~n\\[5pt]
    4 \times \left\lceil \frac{(n+2)}{6} \right\rceil - 2 & \text{for even} ~n
\end{cases}
\end{equation}
The minimum power of $\lambda_3$ as a function of the phonon number $n$ and the power $i$ of $q^2/(2m_d\omega_0)$ for $i>1$ is given by,
\begin{align}\label{eq:cubic_scaling}
\nu(n,i) = \text{max}\Big( 2 \times \left\lceil \frac{|n-i|}{3}\right\rceil~,~2\Big), \quad i > 1.
\end{align}
We show the expansion of the structure factor in the powers of $\lambda_3$ and $q^2/(2m_d\omega_0)$ schematically in Fig.~\ref{fig:anharmonic_schematic}, where we drop the numerical coefficients for all the terms and only illustrate the behavior of the powers of $\lambda_3$ and $q^2/(2m_d\omega_0)$.  In the right part of the schematic, we show the behavior of the $n$-phonon term for $n>3$, and in the left part of the schematic, we show the expansion for $n =$ 1, 2, and 3.

The relationship between the powers in $\lambda_3$ and the powers of $q^2/(2m_d\omega_0)$ in \eqref{eq:cubic_scaling} can also be understood in the following way. The powers of $q^2/(2m_d\omega_0)$ that appear at $\mathcal{O}(\lambda_3^\nu)$ can range from $n-3\nu/2$ to $n+3\nu/2$, with the minimum power allowed being 1, and $\nu$ being an even positive integer. Contributions from powers larger than $n$ are suppressed in the kinematic regime $q \ll \sqrt{2m_d\omega_0}$. But powers smaller than $n$ can lead to significant corrections in the same regime. 

For example, the anharmonic contribution to the 2-phonon structure factor has a leading behavior $\propto~\lambda_3^2 q^2/(2m_d\omega_0)$, which is expected to dominate the harmonic behavior $\propto~q^4/(2m_d\omega_0)^2$ for small enough $q$ (explicitly for $q \lesssim \sqrt{2m_d\omega_0} \lambda_3$). Assuming $m_d \sim 28~\text{GeV}$, $\omega_0 \sim 40~\text{meV}$, and a typical value of $\lambda_3 \sim 0.01$, we expect the anharmonic contribution to start to dominate for $q \lesssim 0.5~\text{keV}$. This kinematic regime does not strictly satisfy the conditions for the incoherent approximation which are assumed in this calculation. However, it is interesting to note here that the size of this anharmonic correction roughly matches onto the result for the 2-phonon structure factor in the long-wavelength limit ($q \ll 1/a$)~\cite{Campbell-Deem:2019hdx,Campbell-Deem:2022fqm}, where it was found that anharmonic interactions give up to an order of magnitude correction to the structure factor. At the edge of the Brillouin Zone $q \sim 2\pi/a \sim{\mathcal{O}}({\rm keV})$, with the typical values used above, we find in the toy model an $O(\sim 25\%$) correction at the boundary of the valid region for the incoherent approximation. 

\begin{figure*}
  \centering \includegraphics[width=0.94\linewidth]{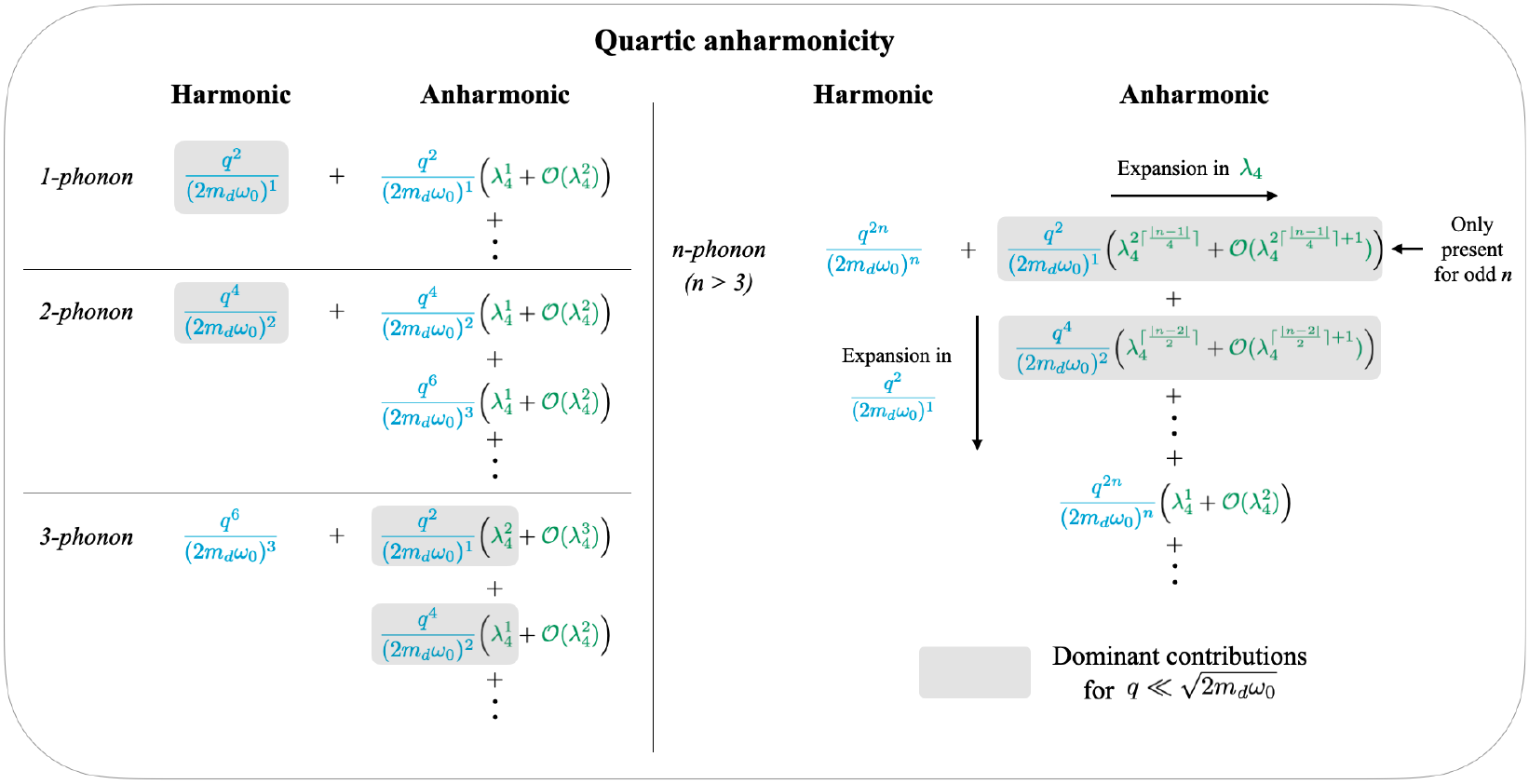}\caption{Expansion of the structure factor in phonon number $n$, $q^2/(2m_d\omega_0)$, and $\lambda_4$ for a quartic perturbation ($k=4$ in \eqref{eq:toy_hamiltonian_perturbation}). The right part shows the general behavior of the $n$-phonon term for $n>3$, while the left part shows the expansion for $n=$1, 2, and 3. Shaded terms show the dominant contributions when $ q\ll \sqrt{2 m_d \omega_0}$, which comes from the anharmonic terms for $ n > 2$. Similar to the above, individual terms might not be present if there is a numerical cancellation in the coefficients. }\label{fig:anharmonic_schematic_quartic}
\end{figure*}

For $k$ equal to 4, which corresponds to a quartic perturbation to the harmonic potential, the calculation proceeds similarly to the cubic case discussed above, except for some key differences. All the coefficients $b_i$ are either real or imaginary based on whether $n$ is even or odd respectively, and hence the anharmonic corrections appear in all orders of $\lambda_4$. We thus have corrections at $\mathcal{O}(\lambda_4)$. For even $n$, coefficients $b_i$ only have even powers of $q$, and thus cannot generate terms $\propto q^2$ in the squared matrix element. The leading behavior for even $n$ is thus $\propto q^4$. For odd $n$ however, the leading behavior is $\propto q^2$, and the minimum power of $\lambda_4$ is given by,
\begin{align}\label{eq:quartic_scaling}
\nu(n,1) =  \text{max}\Big(2 \times \left\lceil \frac{(n-1)}{4}\right\rceil~,~1 \Big).
\end{align}
For powers $i$ greater than 1, the minimum power of $\lambda_4$ for any phonon number $n$ is given by,
\begin{align}
\nu(n,i > 1) = \text{max}\Big(\left\lceil \frac{|n-i|}{2} \right\rceil~,~1\Big).
\end{align}
We show the expansion of the structure factor in the powers of $\lambda_4$ and $q^2/(2m_d\omega_0)$ schematically in Fig.~\ref{fig:anharmonic_schematic_quartic}, where we drop the numerical coefficients for all the terms and only illustrate the behavior of the powers of $\lambda_4$ and $q^2/(2m_d\omega_0)$. Similar to Fig.~\ref{fig:anharmonic_schematic}, we are only illustrating the minimum allowed powers of $\lambda_k$ in perturbation theory for $n>3$. Due to numerical cancellations, the leading $\lambda_k$ power can vanish in some cases.  

\subsubsection{Limitations of perturbation theory}

\begin{figure}[t]
	\centering
\includegraphics[width=\hsize]{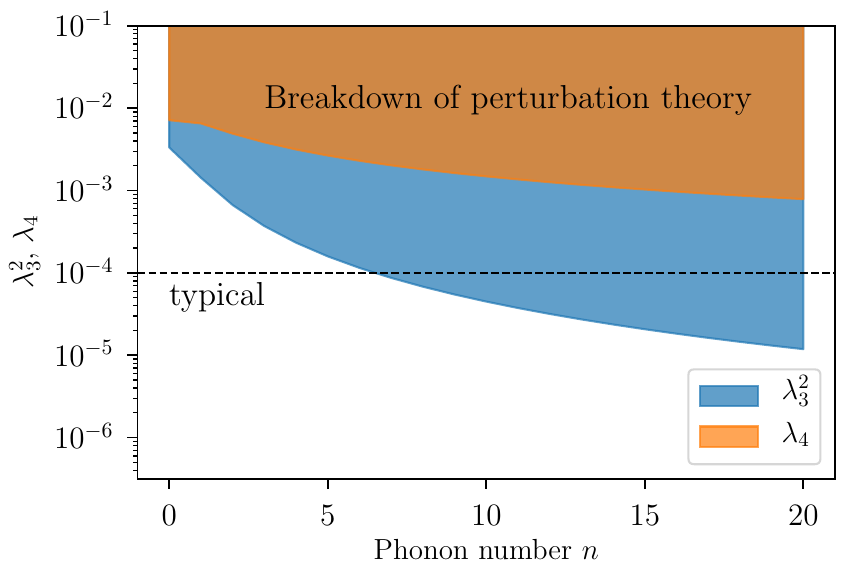}
	\caption{Perturbativity bound on $\lambda_3^2$ and $\lambda_4$ as a function of phonon number $n$. The bound is based on the criteria of \eqref{eq:criteria} that the leading correction to the energy $E_n$ is at most 10\%. The dashed line shows the typical coupling sizes in Si and Ge crystals.}
	\label{fig:nonperturbative_regime}
\end{figure}

Our analysis has focused on the regime $q \ll \sqrt{2m_d\omega_0}$ because this corresponds to a low mean phonon number. For large enough $n$, perturbation theory will start to break down. Equivalently, for a given $n$, perturbation theory will only be valid for $\lambda_k$ sufficiently small. 

For a particular phonon number $n$, if the energy correction in \eqref{eq:energyshiftdef} is of the same order as the unperturbed energy eigenvalue $(n+\frac{1}{2})\omega_0$, the perturbation can no longer be treated as small.  Based on this, we set an upper bound on $|\lambda_k|$ by requiring that
\begin{align}
|E_n' -\Big(n+\frac{1}{2}\Big)\omega_0| \sim 0.1 \times \Big(n+\frac{1}{2}\Big)\omega_0.
\label{eq:criteria}
\end{align}

At leading order, the correction for $k$ equal to 3 (i.e. a cubic perturbation) is given by
\begin{align}
E_n' -\Big(n+\frac{1}{2}\Big)\omega_0 &= \lambda_3^2~\omega_0~(9 n^3 + 9(n+1)^3 \nonumber \\& + (n+3)(n+2)(n+1)\nonumber \\& + n(n-1)(n-2)) + \mathcal{O}(\lambda_3^4).
\end{align}
The equivalent result for $k=4$ reads,
\begin{align}
E_n' -\Big(n+\frac{1}{2}\Big)\omega_0 &= \lambda_4~\omega_0~((n+1)(n+2) + (n+1)^2 \nonumber \\& + 2(n+1)(n+2) + n^2 \nonumber \\& + n(n-1)) + \mathcal{O}(\lambda_4^2).
\end{align}
Using the equations above, we get the critical value of $\lambda_3^2$ and $\lambda_4$ compatible with the perturbation theory expansion. These are shown in Fig.~\ref{fig:nonperturbative_regime}. With the analytic structures of the energy corrections shown above, we see that the perturbativity bound on $\lambda_3^2$ ($\lambda_4$) has a scaling $\propto 1/n^2 (\propto 1/n)$, where $n$ is the phonon number. For typical values of $\lambda_3 \sim 0.01$, we see that the perturbation theory is valid only up to $n \sim 6-7$. Furthermore, perturbation theory is impractical for calculating corrections at small $q$ and very high phonon number $n$, since these corrections will be a very high order in the anharmonicity parameter. 

To deal with these limitations, we consider two different approaches in this paper. Since high $n$ is associated with high $\omega$ and $q$, in the next section we will use the impulse approximation to account for anharmonic effects at high $q$. In Appendix~\ref{sec:morse}, we also study a special anharmonic potential, the Morse potential, where it is possible to obtain exact results. We use this as a case study to validate the perturbation theory and impulse approximation results.

\subsection{Impulse Approximation for $q \gg \sqrt{2 m \omega_0}$}
\label{sec:impulse}

As we have shown, perturbation theory quickly goes out of control beyond the first few number of phonons. Resumming the anharmonic interaction is usually needed for the structure factor when $q$ or $\omega$ is large.
Consider the following phase space 

\begin{align}
    \label{eq:impulse_regime}
    \textrm{Impulse regime:}&\quad q \gg \sqrt{2 m \omega_0}, \\
    &\quad \omega \sim \frac{q^2}{2m}+\mathcal{O}(\sqrt{\omega {\omega}_0}),\nn
\end{align}
It has previously been shown \cite{Campbell-Deem:2022fqm,gunnwarner} in the harmonic case, that one can calculate the structure factor by using a saddle point approximation in the time-integral representation of the structure factor.
This is called the ``impulse approximation" since the steepest-descent contour is dominated by small times, which can be interpreted physically as an impulse.

We begin with the structure factor in Eq.~\eqref{eq:structure_factor_toy_1D}, which can be decomposed as contributions from each atom $d$, $S_{\text{toy}}(q,\omega) = \sum_d n_d |f_d|^2 S_{\text{toy},d}(q,\omega)$. Then we rewrite the energy conservation delta function as a time integral
\begin{align}
    &S_{\text{toy},d}(q,\omega) \nonumber \\
    &\equiv \sum_f\int dt\, e^{i(E_f-E_0-\omega)t} \left|    \mel{\Phi_f}{e^{iqx}}{\Phi_0} \right|^2 \nonumber \\
    &= 
    \int dt\, e^{-i\omega t}
    \sum_f \mel{\Phi_0}{e^{-iqx}}{\Phi_f}
    \mel{\Phi_f}{e^{iHt}e^{iqx}e^{-iHt}}{\Phi_0} \nonumber \\
    &=
    \int dt\, e^{-i\omega t}
    \mel{\Phi_0}{e^{-iqx}e^{iqx(t)}}{\Phi_0},
    \label{eq:S_time}
\end{align}
where in the second equality we use the fact that $|\Phi_0\rangle$ and $|\Phi_f\rangle$ are eigenfunctions of $H$, and in the third equality we use the completeness relation and the time-dependent position operator $x(t)= e^{iHt}xe^{-iHt}$.
The final expression is the well-known structure factor in the time domain.
Using the above representation of the structure factor,
\begin{align}
    S_{\text{toy},d}(q,\omega) &= \int_{-\infty}^{\infty} dt ~ \langle e^{-iqx}e^{iqx(t)}\rangle~e^{-i\omega t} \nn \\
    &=  \int_{-\infty}^\infty dt ~\langle e^{-iqx} e^{iHt} e^{iqx} \rangle ~ e^{-i (E_0 + \omega) t},
    \label{eq:S_time_simple}
\end{align}
We can further simplify this using the fact that $e^{iqx}$ acts as a translation operator on momentum $p$, $e^{-iqx} \,p\, e^{iqx} = p+q$. Applying the translation on the full Hamiltonian yields
\begin{align}
    e^{-iqx} H(x,p) e^{iqx}=H(x,p+q).
\end{align}

Here we generalize the impulse approximation to any 1D Hamiltonian, $H(x,p)=\frac{p^2}{2m}+V(x)$, which satisfies
\begin{align}
    H(x,p+q) = H(x,p) + \frac{q^2}{2m}+ \frac{q}{m} p. 
    \label{eq:impulse_condition}
\end{align}
One can also generalize impulse approximation to a generic potential $V(x,p)$ as long as the above holds in the limit of large $q$.\footnote{In this case, the impluse regime in Eq.~\eqref{eq:impulse_regime} needs to be replaced as $\omega \sim \frac{q^2}{2m}+ \frac{q}{m}\langle p\rangle$ and we impose Eq.~\eqref{eq:impulse_condition} holds up to $\mathcal{O}\left(\omega^2_0/{q} \right)$ correction.}
In other words, we require that the Hamiltonian in the large momentum limit is dominated by the kinetic energy $\frac{p^2}{2m}$, not the potential.
We can then obtain reliable theoretical predictions in the impulse regime even with large number of phonons.

Applying the above to Eq.~\eqref{eq:S_time_simple}, the structure function now reads
\begin{align}
    S_{\text{toy},d}&(q,\omega) 
    = \int_{-\infty}^\infty dt \, \left \langle e^{i H(x,p+q) t} \right \rangle \, e^{-i (E_0 + \omega)t} \nn \\
    &\approx \int_{-\infty}^\infty dt \, \left\langle e^{i \left(H+\frac{qp}{m} \right) t} \right\rangle e^{-i \left(E_0 + \omega-\frac{q^2}{2m} \right) t},
\end{align}
where we translate the momentum in the first line and use Eq.~\eqref{eq:impulse_condition} in the second line. Note that $H = H(x,p)$ throughout and we drop the argument for brevity. The last line is exact for potentials that depend only on $x$.

Now we can apply the saddle point approximation to evaluate the time integral.
Defining $H' \equiv H + \frac{pq}{m}$, we can write
\begin{align}
    S_{\text{toy},d}(q,\omega) &= \int_{-\infty}^\infty dt \, e^{f(t)},
\end{align}
where
\begin{align}
    \label{eq:impulse_exponent}
    f(t) &\equiv \ln \langle e^{i H' t} \rangle - i t 
\left (E_0 + \omega - \frac{q^2}{2m} \right).
\end{align}

In order to calculate this object, we can expand $\ln \langle e^{i H' t} \rangle$ in small $t$. The first few terms in this expansion are given by
\begin{align}
    \nonumber
    f(0) &= 0 \\
    f'(0) &= i \Big( \frac{q^2}{2m} - \omega \Big)
    \nonumber
    \\
    \nonumber
    f''(0) &= i^2 \Big( \langle H'^2 \rangle - \langle H' \rangle^2 \Big) \\ \nonumber &= -\frac{q^2}{m^2} \Big( \langle p^2 \rangle - \langle p \rangle^2 \Big)
    \\
    \nonumber
    f^{(3)}(0) &= i^3 \Big( \langle H'^3 \rangle -3 \langle H' \rangle \langle H'^2 \rangle + 2 \langle H' \rangle^3 \Big) \\ \nonumber &= -i \Big( \frac{q^2}{m^2} \langle p [H,p] \rangle + \frac{q^3}{m^3} \langle p^3 \rangle \Big)
    \\
    \nonumber
    f^{(4)}(0) &= i^4 \Big ( -6 \langle H' \rangle^4+ 12 \langle H'\rangle^2 \langle H'^2 \rangle^2
    \\
    \nonumber
    & \qquad - 3 \langle H'^2 \rangle^2 - 4 \langle H' \rangle \langle H'^3 \rangle + \langle H'^4 \rangle \Big) \\ \nonumber &= -\frac{q^2}{m^2} \langle [p, H]^2 \rangle + \frac{q^3}{m^3} \langle [[p,H], p^2] \rangle
    \\
    &\qquad + \frac{q^4}{m^4} \Big( \langle p^4 \rangle - 3 \langle p^2 \rangle^2 \Big)
    \label{eq:impulse_fourth_order}
    \\
    \nonumber
    & \ldots
\end{align}
In the harmonic approximation, only the terms proportional to $q^2$ are nonzero. As a result, only the first few expansion terms are needed as long as $t \ll \frac{1}{\omega_0}$ since $f^{(n+1)}/f^{(n)}$ is of order $\omega_0$. Then one can solve for the  saddle point $t_I$ by solving $f'(t_I) \approx f'(0) + f''(0) t_I = 0$, which gives
\begin{equation}
    i t_I = \frac{m^2 (\omega - \frac{q^2}{2m})}{q^2 \sigma_p^2},
    \label{eq:tI_secondorder}
\end{equation}
where
\begin{align}
    \sigma^2_p \equiv \langle p^2\rangle-\langle p\rangle^2 = \langle p^2\rangle.
\end{align}
In the last equality we use the fact that $\langle p\rangle=0$ for a $V(x)$ potential since $\langle p\rangle \propto \langle [x,H]\rangle=0$. Although $t_I$ is formally imaginary, its magnitude is small and close to the origin in the impulse regime. Since there is no pole around this saddle point, we can approximate the time integral by the saddle point and find
\begin{align}
    S_{\text{toy},d}(q,\omega) &\approx
    \sqrt{ \frac{2 \pi}{-f''(t_I)}} \,e^{f(t_I)} \nn \\
    &= \frac{\sqrt{2\pi} m}{q \sigma_p}
    \exp\left(-\frac{m^2(\omega-\frac{q^2}{2m})^2}{2 q^2 \sigma_p^2}\right).
    \label{eq:impulse_result}
\end{align}
For large energy depositions the Gaussian becomes narrowly peaked around $\omega = q^2/2m$, and this reproduces the nuclear recoil limit~\cite{Campbell-Deem:2022fqm}.

In the presence of anharmonic interactions, other powers of $q$ will be present in the expansion of \eqref{eq:impulse_fourth_order}. In general, the $f^{(n)}$ term will have a $q^n$ term with coefficient of $\mathcal{O}(\lambda).$ In this case, $f^{(n+1)}/f^{(n)} \sim q \sqrt{\omega_0/m}$. Higher orders will then be important in the expansion of $f(t)$ for sufficiently large $q$ or $t$.  For a given $q$, the higher order corrections become relevant for $|t| \gtrsim \sqrt{m/\omega_0}/q \sim 1/\sqrt{\omega \omega_0}$ in the impulse regime. Including these corrections is difficult in general, but we can continue to use the second order expansion giving \eqref{eq:impulse_result} as long as $|t| \lesssim 1/\sqrt{\omega \omega_0}.$  According to \eqref{eq:tI_secondorder}, this corresponds to a condition on how close $\omega$ is to $q^2/(2m)$. Since $q^2 \sim 2m \omega$ and $\sigma_p^2 \sim m \omega_0,$ this implies that
\begin{equation}
    \vert t_I \vert \sim \frac{\vert \omega - \frac{q^2}{2m} \vert}{\omega \omega_0} \rightarrow \vert \omega - \frac{q^2}{2m} \vert \lesssim \sqrt{\omega \omega_0}.
    \label{eq:saddle_point_condition}
\end{equation}
We see that the distance of $\omega$ from $\frac{q^2}{2m}$ sets the size of $t_I,$ which in turn tells us the regime for the validity for the approximation \eqref{eq:impulse_result}. The condition \eqref{eq:saddle_point_condition} is approximately the same condition that $\omega$ is within the Gaussian width in \eqref{eq:impulse_result}, and keeping terms in $f(t)$ only up to $f''(0)$ is self-consistent near $\omega = \frac{q^2}{2m}$.

Therefore, in the presence of anharmonic interactions, the above structure factor result \eqref{eq:impulse_result} remains valid in the impulse regime~\eqref{eq:impulse_regime}. The only modification is in $\sigma_p^2$. Considering perturbations in $V(x)$ up to $x^4$ and recalling that the expectation value is with respect to the full ground state, we find that
\begin{equation}
    \sigma_p^2 = \langle p^2 \rangle = \frac{m \omega_0}{2} \Big(1 - 44 \lambda_3^2 + 12 \lambda_4 + \cdots \Big)
\end{equation}
at leading order in $\lambda_3, \lambda_4.$ The nuclear recoil limit is again reproduced, with a small modification to the width of the Gaussian envelope due to anharmonic couplings.  Note that in order to calculate the structure factor far from $\omega = \frac{q^2}{2m},$ we must include additional orders in $f(t)$ and $t_I$. We do not perform these higher order calculations for the final results in this paper since they have a negligible effect on the integrated rates, but we provide the procedure for completeness in App.~\ref{sec:impulse_appendix}.

Finally, we approximate the effect that introducing the full crystal lattice has on this single atom result. Up until the evaluation of various moments of $H',$ the impulse approximation is fully model-independent. We just have to make an adjustment to the final evaluation of $\langle p^2 \rangle$. The states in the full crystal theory are smeared by the phonon density of states, so we calculate $\langle p^2 \rangle$ via the following prescription
\begin{align}
    \langle p^2 \rangle &= \frac{m \omega_0}{2} \Big( 1 + g(\lambda) \Big)\nonumber \\
    &\xrightarrow{\mathrm{crystal}} \int d \omega' D( \omega') \frac{m \omega'}{2} \Big( 1 + g(\lambda) \Big),
    \label{eq:psquared_prescription}
\end{align}
where $g(\lambda)$ is the anharmonic correction calculated in the single-atom potential.  Essentially, we have used the average single phonon energy to calculate $\langle p^2 \rangle.$ In the harmonic limit, \eqref{eq:impulse_result} then exactly matches the impulse result from \cite{Campbell-Deem:2022fqm}.

In summary, in this section we have demonstrated the general behavior of anharmonic effects with $q$ and $\omega$. We have shown that they are indeed negligible at high $q$ and $\omega \sim q^2/2m_d$, consistent with the intuition that scattering can be described by elastic recoils of a free nucleus. The effects grow for $q \ll \sqrt{2 m_d \omega_0}$ and at low $q$ they may dominate the structure factor. This roughly matches onto the results of Refs.~\cite{Campbell-Deem:2019hdx,Campbell-Deem:2022fqm}, which found that for $q < 2 \pi/a$ anharmonic effects can have a large impact on the two-phonon rate .

\section{Numerical results for 1D anharmonic oscillator}\label{sec:exact}

Having demonstrated the analytic behavior of the dynamic structure factor in the previous section, we now turn to obtaining numerical results using realistic potentials. We will perform concrete calculations for Si and Ge as representative materials while briefly commenting on others. As discussed in Sec.~\ref{sec:anharmonic_properties}, we adopt an empirical model of interatomic interactions that encodes the anharmonicity in the potential. We use this empirical model to calculate a single atom potential, which we then use to evaluate the structure factor numerically. 

As stated in Sec.~\ref{sec:toy_anharmonic}, we start by fitting the single atom potential in a particular direction onto a 1D potential of the form,
\begin{align}
    \label{eq:1D_potential}
    V_d(x) &= \frac{1}{2} m_d \omega_0^2 x^2 + \lambda_3 \omega_0 (\sqrt{2 m_d \omega_0} x)^3  \nonumber \\
    &\quad + \lambda_4 \omega_0(\sqrt{2 m_d \omega_0} x)^4.
\end{align} In the fit, $\omega_0, \lambda_3, \lambda_4$ are free parameters but in order to reproduce the harmonic limit, we then make the replacement $\omega_0 = 1/\overline{\omega^{-1}}$, which is calculated from the phonon density of states and gives a slightly different numerical value. This is motivated by the harmonic case discussed in Sec.~\ref{sec:structure_factor_harmonicToToy}. 
We do not consider anharmonic terms $\propto x^k$ for $k \geq 5$ as we observe that the anharmonic potential along any direction is dominated by the cubic and the quartic terms. 

We find that the maximum anharmonicity is typically along the nearest neighbor direction $(x,y,z)=(1,1,1)$. For computing results, we will consider the potential along this direction, which represents maximum anharmonicity, as well as the potential in an orthogonal direction $(x, y, z) = (1, -2, 1)$, which represents an intermediate value for the anharmonicity. Using the aforementioned interatomic models, we find anharmonicity strengths ranging from $\lambda_3 \sim 6 \times 10^{-3}$ to $10^{-2}$ and $\lambda_4 \sim (2-3) \times 10^{-4}$. For Si and Ge, the results are same for either atom in the unit cell.

\begin{figure*}
\centering
\includegraphics[width=0.49\linewidth]{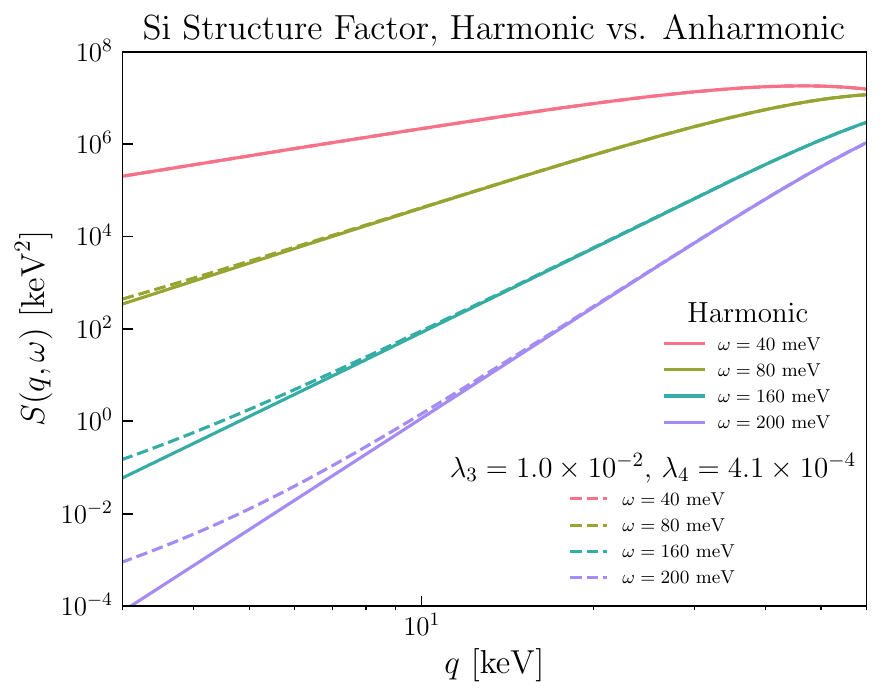}
\includegraphics[width=0.49\linewidth]{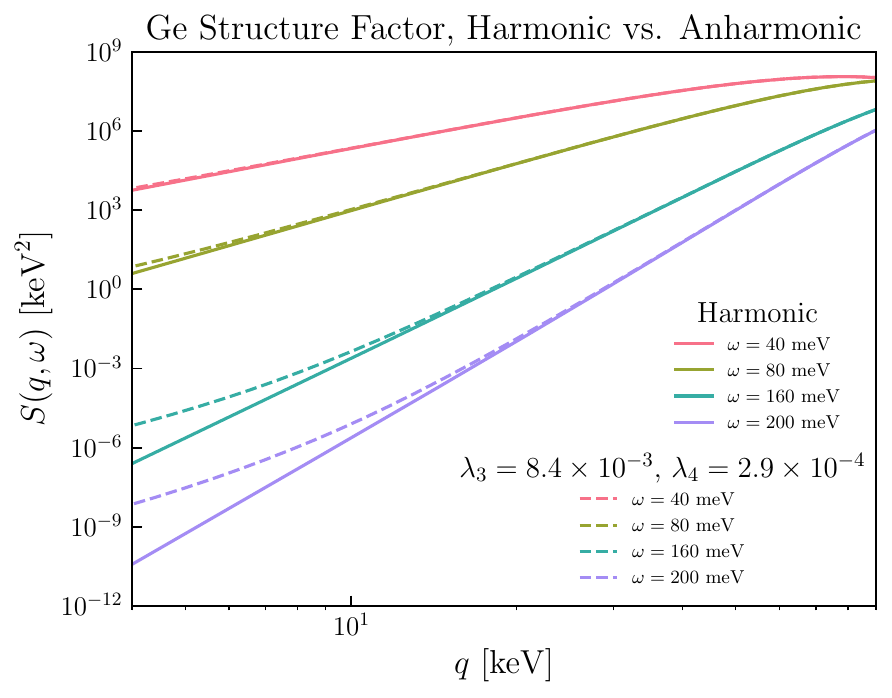}
\vspace{0.5cm}
\caption{ 
\label{fig:si_structure_factor_q} {\bf $q$-dependence of structure factor:} We compare the structure factor in the harmonic and anharmonic cases, where in the latter case the structure factor is calculated numerically with the maximal anharmonicity. The lines from top to bottom show the structure factor at different $\omega$, corresponding to an increasing minimum phonon number $n$. There are large corrections for $q \ll \sqrt{2 m_d \omega_0}$ when anharmonic interactions are included (dashed), and the corrections become more significant as the threshold is increased. For $q \gg \sqrt{2 m_d \omega_0}$, both cases converge to the same result. For Si, we have $\sqrt{2 m_d \omega_0} \approx 40$ keV while for Ge, $\sqrt{2 m_d \omega_0} \approx 50$ keV. For other materials, this quantity is listed in Table~\ref{fig:material_properties}. The incoherent approximation momentum cutoff is $q_\mathrm{BZ} < 2 \pi / a \sim 2.2$ keV for both crystals. \vspace{1cm}
}
\end{figure*}

\begin{figure*}
\centering
\includegraphics[width=0.98\linewidth]{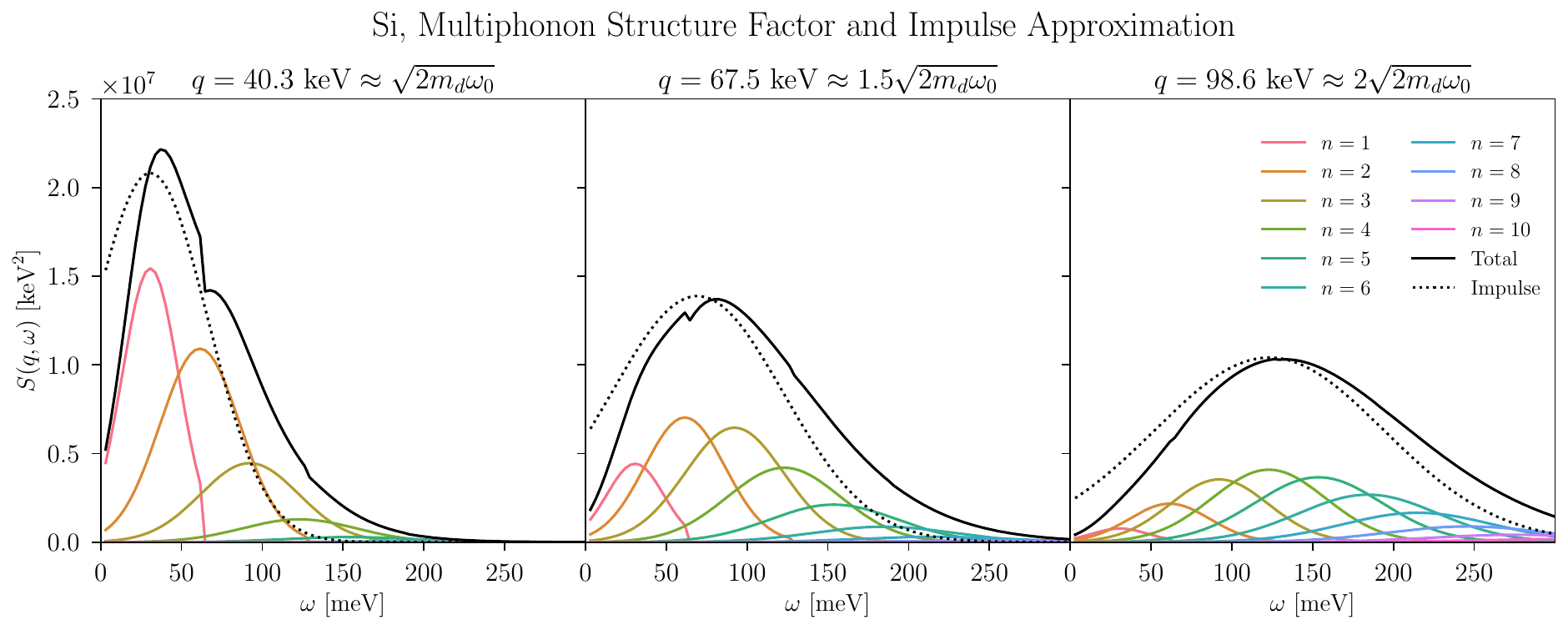}  
\vspace{0.5cm}
\caption{ 
\label{fig:impulse_approx} {\bf $\omega$-dependence of structure factor:}  For different $q$ values, we show the decomposition of the structure factor into individual $n$ phonon terms, where the energy-conserving delta function has been smeared as in \eqref{eq:structure_factor_final}. Note that the maximum anharmonicity has been included in the numerical calculation, but the result is nearly identical to the harmonic result for these $q$ values, as shown in Fig.~\ref{fig:si_structure_factor_q}. The dotted line shows the impulse approximation, which starts to become a good approximation as $q$ increases above $\sqrt{2 m_d \omega_0}$. 
}
\end{figure*}

Given the 1D potential in \eqref{eq:1D_potential}, we find exact solutions of the 1D eigenvalue and eigenvector problem using a simple finite difference method. We take a first order discretization of the Laplace operator and solve the discretized time-independent Schr\"odinger equation in a box. The box grid interval size must be small enough to resolve the maximum momentum scales of interest, which in this case depends on the highest excited state needed in the calculation. Also, the minimum box size required depends on the spatial extent of the highest excited state used. As seen in Sec.~\ref{sec:impulse}, the impulse approximation suffices for $q > \mathcal{O} (\mathrm{few}) \times \sqrt{2 m_d \omega_0}$. Beyond this momentum, we no longer need to calculate excited states since the structure factor in the impulse limit is independent of the details of the highly excited states. The \emph{n}th excited state is most relevant at momenta $q \sim \sqrt{n} \sqrt{2 m \omega_0}.$ Therefore, to complete our calculation below the impulse limit, we include the first 10 excited states. The results for these eigenstates are converged above a box size of $ \sim 10/\sqrt{2 m \omega_0}$ and grid size of $ \sim 0.1/\sqrt{2 m \omega_0}$.

We now use these numerical eigenstates and energies to calculate the structure factor in Eq.~\eqref{eq:structure_factor_toy_1D}. We apply a prescription for the energy-conserving delta function similar to that used in the harmonic 1D oscillator, Eq.~\eqref{eq:replacement_prescription}. The final result at momenta below the impulse regime ($q < 2 \sqrt{2 m \omega_0}$) is,
\begin{align}
    \nonumber
    S(q,& \omega) = 2\pi \sum_{d} n_d ~|f_d|^2 \sum_f \left|    \mel{\Phi_f}{e^{iqx}}{\Phi_0} \right|^2 \nonumber \\& \times \frac{1}{\sqrt{2\pi f(n) \sigma^2}} e^{- \frac{(\omega - f(n) \omega_0)^2}{2 f(n) \sigma^2}} \times \Theta(\omega_{\mathrm{max}} - \omega),
    \label{eq:structure_factor_final}
\end{align}
where 
\begin{align}
    \omega_0 &= \Big( \int d \omega \omega^{-1} D(\omega) \Big)^{-1},
    \\
    \label{eq:single_ph_width}
    \sigma & = \sqrt{\frac{\int d\omega \omega D(\omega)}{\omega_0} - \frac{1}{\omega_0^2}},
    \\
    \omega_\mathrm{max} &= f(n) \times (\mathrm{min}(\omega) \vert D(\omega) = 0 )
\end{align}
and $f(n), \vert \Phi_0 \rangle, \vert \Phi_f \rangle$ are given by the numerically solved eigenenergies and eigenstates, respectively. $D (\omega)$ is the single phonon density of states calculated with DFT \cite{Jain2013}. In this work we assume equal couplings of DM with all nucleons so that $f_d = A_d$, where $A_d$ is the atomic mass number. In the equations above, we have included a sum over all atoms in the unit cell $d$ with density $n_d$, and in general the atomic potentials and density states can also depend on $d$, although for Si and Ge we do not include this.

In the impulse regime ($q > 2 \sqrt{2 m \omega_0}$), we have shown in Sec.~\ref{sec:impulse} that the structure factor for any position-dependent potential is approximated by a Gaussian envelope,
\begin{equation}
\label{eq:structure_factor_impulse}
S(q, \omega) \approx \sum_{d} n_d |f_d|^2 \sqrt{\frac{2 \pi}{\frac{q^2}{m^2} \langle p^2 \rangle}} e^{- \frac{ \big( \omega - \frac{q^2}{2m} \big)^2}{2 \frac{q^2}{m^2} \langle p^2 \rangle}},
\end{equation}
where the the expectation values are all computed in the ground state and adjusted to the average single phonon energy via \eqref{eq:psquared_prescription}. Now we simply use the numerical ground state of the anharmonic potential \eqref{eq:1D_potential} to calculate $\langle p^2 \rangle$ and therefore obtain the structure factor. Note that the anharmonic contribution is essentially negligible in the impulse limit, since corrections to $\langle p^2 \rangle$ are $\propto$ $\lambda_3^2, \lambda_4$.

Fig.~\ref{fig:si_structure_factor_q}-\ref{fig:impulse_approx} shows numerical results on the structure factor for Si and Ge, taking the maximum anharmonicity in either case. In Fig.~\ref{fig:si_structure_factor_q}, the structure factor as a function of $q$ is shown. As $\omega$ (and therefore minimum phonon number~$n$) is increased, there is a larger anharmonic correction at small $q$.  
This can be understood by looking at the $q$ scalings discussed in Sec.~\ref{sec:perturbation_theory} and illustrated in Fig.~\ref{fig:anharmonic_schematic} and Fig.~\ref{fig:anharmonic_schematic_quartic}. At low $q$ and thus DM mass, the contributions from the anharmonic structure factor can give smaller powers of $\frac{q^2}{2 m_d \omega_0}$ compared to the leading harmonic term $\Big( \frac{q^2}{2 m_d \omega_0} \Big)^n$, so the enhancement grows with $n$.
At high $q$, results converge to the harmonic result, consistent with our discussion of the impulse regime in Sec.~\ref{sec:impulse}. We see this also in Fig.~\ref{fig:impulse_approx}, which shows the structure factor at different $q$. The impulse approximation becomes better as $q \gg \sqrt{2 m_d \omega_0}$, and is indistinguishable from the harmonic case.

\subsection{Impact on DM scattering rates}\label{sec:scattering_rates}

\begin{figure*}
\centering
\includegraphics[width=0.47\linewidth]{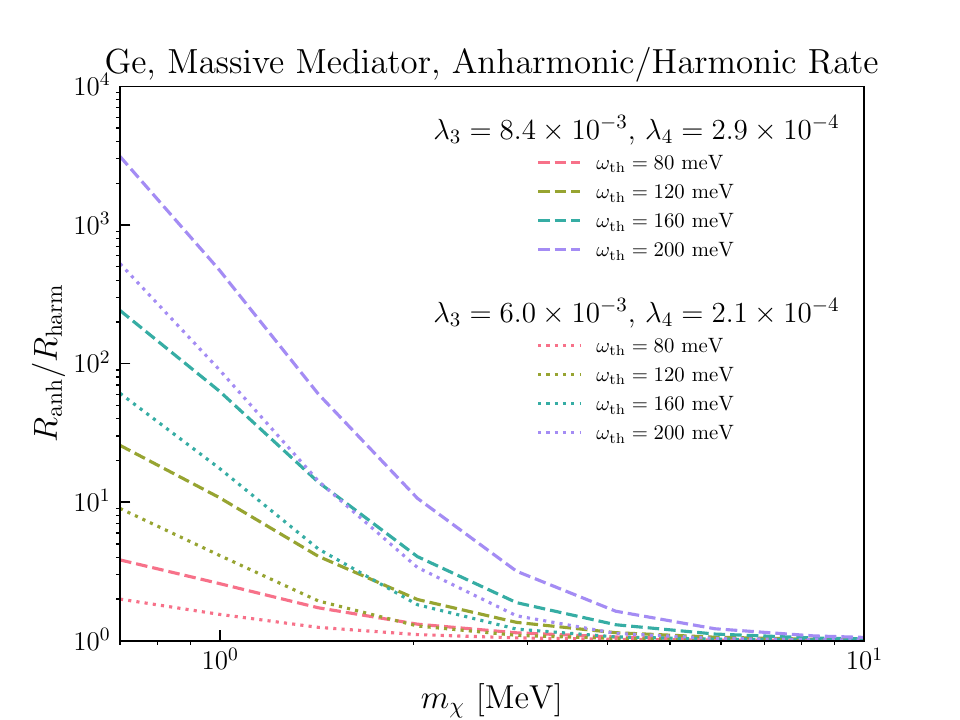} 
\includegraphics[width=0.47\linewidth]{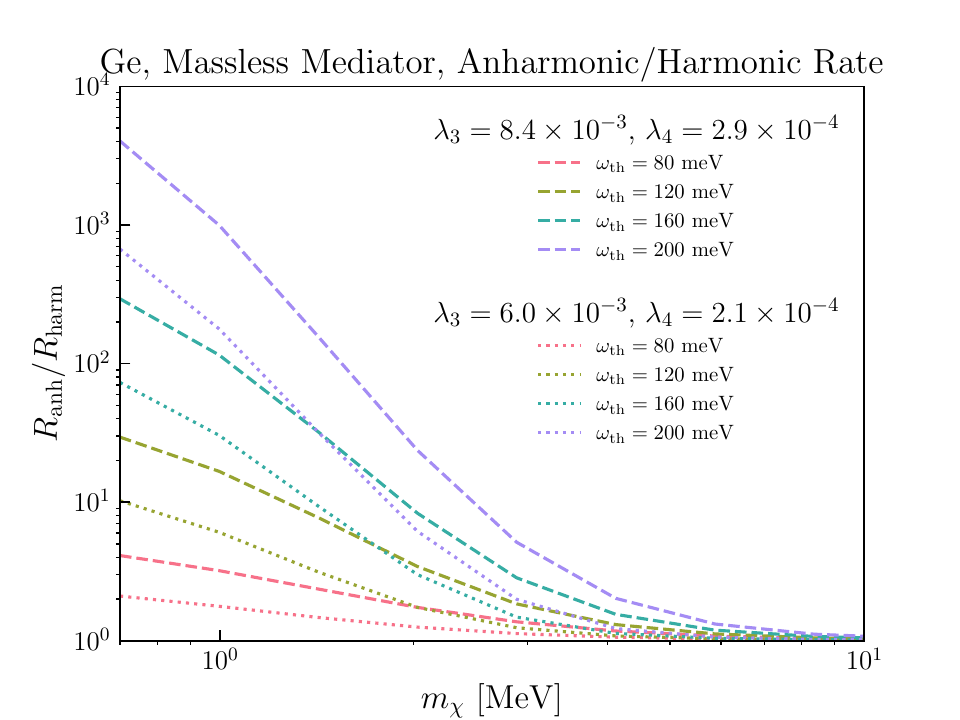} 
\includegraphics[width=0.47\linewidth]{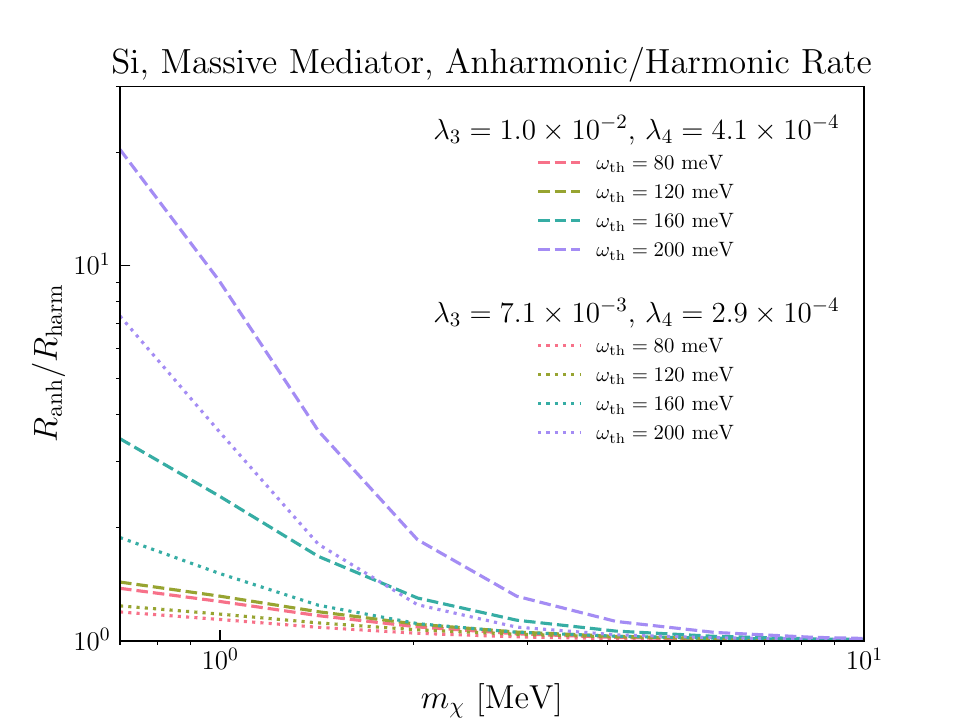} 
\includegraphics[width=0.47\linewidth]{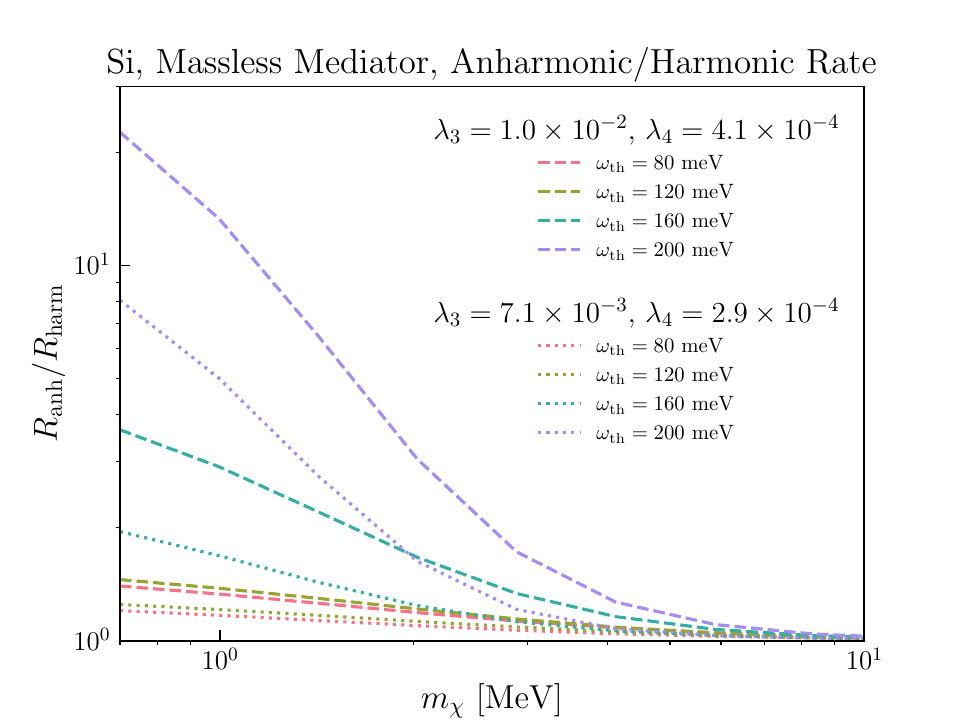} 
\caption{ 
\label{fig:anharmonic_ratios} \textbf{Ratio of anharmonic to harmonic rate.} For each material (Ge and Si) we consider two representative values of the anharmonic couplings. The larger set corresponds to a direction of maximal anharmonicity while the other set corresponds to an orthogonal direction of intermediate anharmonicity. Anharmonic effects become more important for DM masses near the MeV scale and for larger energy thresholds.}
\end{figure*}

\begin{figure*}
\centering
\includegraphics[width=0.47\linewidth]{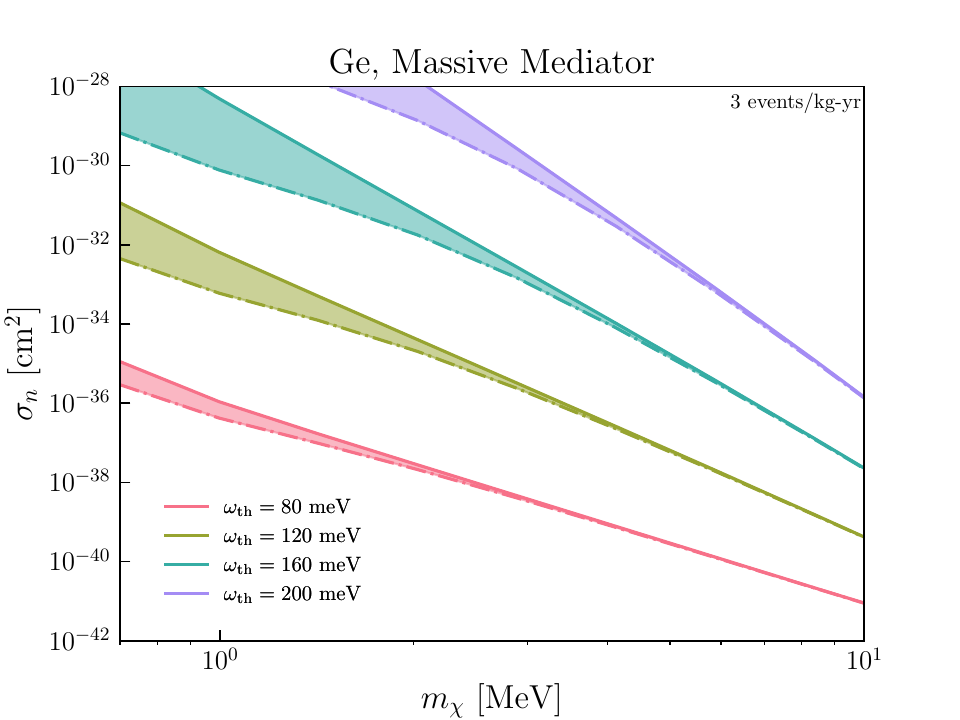} 
\includegraphics[width=0.47\linewidth]{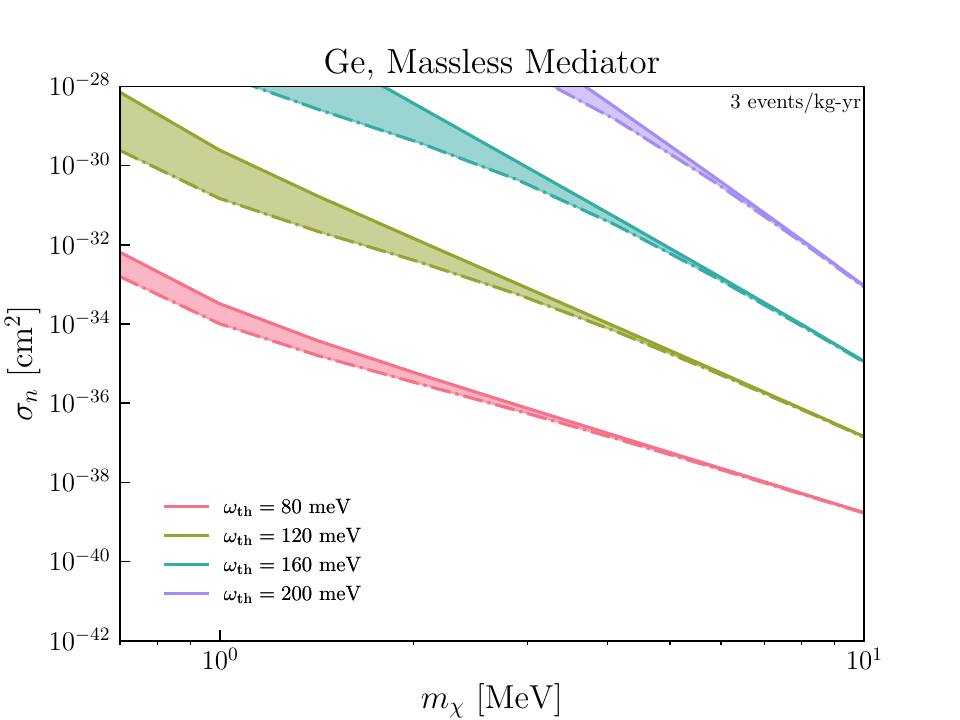} 
\includegraphics[width=0.47\linewidth]{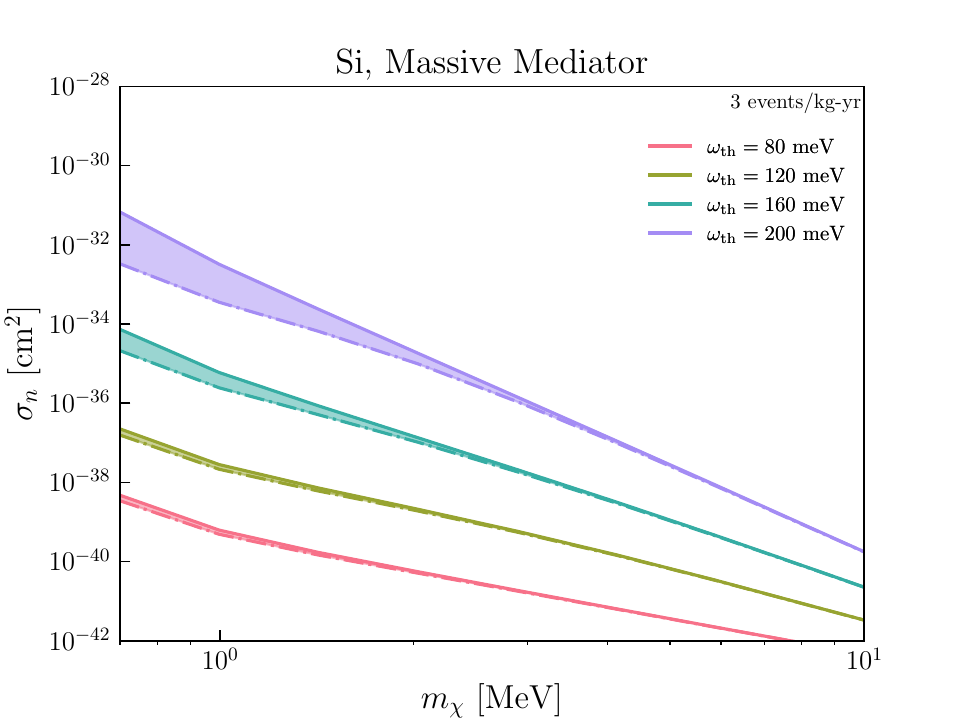} 
\includegraphics[width=0.47\linewidth]{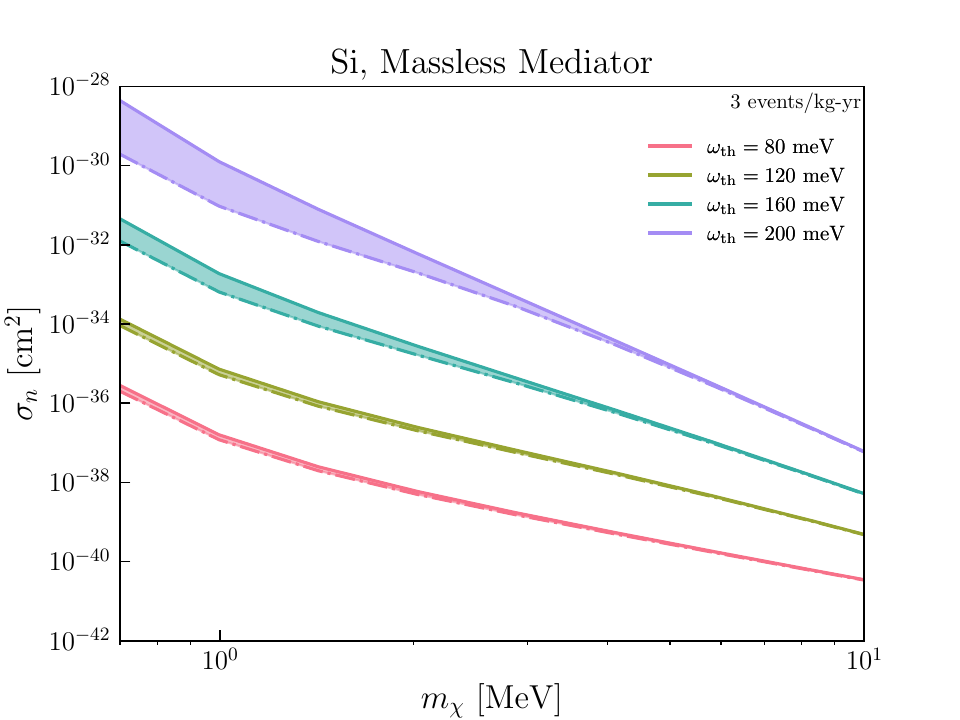} 
\caption{ 
\label{fig:cross_sections} {\bf Cross section uncertainty.} Comparison of the cross section corresponding to 3 events/kg-yr in the harmonic (solid) and anharmonic (dot-dashed) cases. The anharmonic result is shown for maximal anharmonicity, and so the shaded band represents our estimate of the theoretical uncertainty due to anharmonic effects. The effects are primarily important for high thresholds and low DM masses, corresponding to large $\sigma_n$, which is generally in tension with existing astrophysical or terrestrial constraints.}
\end{figure*}

We now use the numerical results for the structure factor to compute the DM scattering rates for a range of DM masses and experimental thresholds. Our results are summarized in Figs.~\ref{fig:anharmonic_ratios}-\ref{fig:cross_sections}. We consider DM masses in the range $\sim 1-10 $ MeV. The lower end of the mass range is chosen such that the momentum transfers are large enough to satisfy the condition for the incoherent approximation (i.e. $q > 2\pi/a$), while at the upper end of masses it is expected that scattering is described by the impulse approximation~\cite{Campbell-Deem:2022fqm}. It is precisely this mass range where details of multiphonon production are important. We will also consider the two cases of scattering through heavy and light mediators. The goal will be to identify the region of parameter space where the anharmonic effects on the dynamic structure factor affect the scattering rates the most.

In the isotropic limit, the observed DM event rate per unit mass is given by \cite{Campbell-Deem:2022fqm}
\begin{equation}
    \label{eq:rate-isotropic}
    R =  \frac{1}{4\pi \rho_T} \frac{\rho_\chi}{m_\chi}  \frac{\sigma_p}{\mu_\chi^2}  \int \! d^3 \bfv\, \frac{f(\bfv)}{v} \int\displaylimits_{q_-}^{q_+} \! dq \,  \int\displaylimits_{\omega_\mathrm{th}}^{\omega_+} \! d\omega \,   q \, |\Tilde{F}(q)|^2 S(q, \omega),
\end{equation}
where $\rho_\chi$ is the DM energy density, $\rho_T$ is the mass density of the target material, $m_\chi$ is the DM mass, $\mu_\chi$ is the DM-nucleon reduced mass, $\sigma_p$ is the DM-nucleon cross section, and $f(\mathbf{v})$ is the DM velocity distribution. The structure factor $S(q, \omega)$ is given by our numerical results \eqref{eq:structure_factor_final}-\eqref{eq:structure_factor_impulse} and the integration bounds are determined by the kinematically allowed phase space
\begin{align}
q_\pm &\equiv m_\chi v \left(1\pm\sqrt{1-\frac{2\omega_\mathrm{th}}{m_\chi v^2}}\right)\label{eq:qphasespace},\\
\omega_+ &\equiv qv -\frac{q^2}{2 m_\chi}, \label{eq:omegaphasespace}
\end{align}
where the energy threshold of the experiment is denoted by $\omega_\mathrm{th}$.
The $q$-dependence of the DM-nucleus interaction can be encapsulated in the DM form factor $\Tilde{F}(q)$, where $\Tilde{F}(q)=1$ indicates an interaction through a heavy mediator, and $\Tilde{F}(q)=q_0^2/q^2$ indicates an interaction through a light mediator for a reference momentum transfer of $q_0$. 

Note that in general, the strength of the anharmonicity varies with the direction of the recoil of the nucleus, and the structure factor will depend on the direction of the momentum transfer. For simplicity, we are assuming that the anharmonicity strength is uniform in all directions. Our estimate with the maximum anharmonicity thus provides an upper bound on the anharmonic effects on DM scattering.

The DM mass sets the typical momentum-transfer scale $q$ of the scattering, and the experimental energy threshold $\omega_{\text{th}}$ sets the phonon number $n$. Hence, to identify the DM masses and experimental thresholds  where anharmonic effects start to become important, we first need to understand the $q$-values where the anharmonic corrections are large for a particular phonon number $n$. We can estimate this using the perturbation theory results in Sec.~\ref{sec:perturbation_theory}. Note that in our numerical calculation, we find that $\lambda_3$ generally provides the larger anharmonic contribution, so we will focus on a purely cubic perturbation in this discussion. 

For the analysis of a cubic perturbation discussed in Sec.~\ref{sec:perturbation_theory}, we showed that anharmonic effects introduced additional terms to the $n$-phonon structure factor of the form $\propto$ $\lambda_3^{\nu(n,i)} \Big(\frac{q^2}{2 m_d \omega_0}\Big)^i$, see \eqref{eq:perturbation_expansion}. Therefore when $q$ is lower than the scale
\begin{equation}
    \label{eq:anharmonic_q}
    q \lesssim \sqrt{2 m_d \omega_0} \lambda_3^{\nu(n, i)/(2(n-i))},
\end{equation}
terms in the anharmonic structure factor can be of comparable size to the harmonic structure factor. In order to find the largest $q$-scale where the anharmonic contribution starts to become relevant, we can evaluate \eqref{eq:anharmonic_q} for all positive $i<n$, and find the minimum possible exponent of $\lambda_3$. For $n = 2$ or $3$,  the minimum exponent is achieved for $i=1$, for which $\nu(n, 1) = 2$. This gives a $q$-scaling of $q \sim  \sqrt{2 m_d \omega_0} \lambda_3^{1/(n-i)}.$ This tells us that for the 2-phonon case, the anharmonic contribution should begin to become important at $q \sim \sqrt{2 m_d \omega_0} \lambda_3,$ while for the 3-phonon case, the anharmonic contribution becomes important at $q \sim \sqrt{2 m_d \omega_0} \lambda_3^{1/2}$. For a larger number of phonons, this scaling is approximately $q \sim \sqrt{2 m_d \omega_0} \lambda_3^{1/3}$. So we see that higher energy excitations have more significant anharmonic contributions at larger momentum transfers. Below the $q$-scale identified above, the anharmonic contributions are expected to increase substantially with decreasing $q$, as terms $\propto q^{2i}$ for $i<n$ dominate the harmonic scaling $\propto q^{2n}$.

We now recast our analysis concretely in terms of DM mass and experimental energy thresholds as follows. For both massive and massless mediators, the event rate for $n \geq 2$ phonons is always dominated by the large $q$ portion of phase space and energy depositions near the threshold. Therefore the enhancement in the rate due to the anharmonicity roughly corresponds to the enhancement in structure factor evaluated at $S(q=2 m_\chi v, \omega=\omega_\mathrm{th})$, where $v$ is the DM velocity. 
Inserting $q = 2 m_\chi v$ into the condition in \eqref{eq:anharmonic_q} gives a condition on the DM mass:
\begin{equation}
\label{eq:dm_mass_threshold}
m_\chi \lesssim 
\begin{cases}
    \frac{\sqrt{2 m_d \omega_0} \lambda_3}{2 \times 10^{-3}} & n = 2\\
    \frac{\sqrt{2 m_d \omega_0} \lambda_3^{1/2}}{2 \times 10^{-3}} & n = 3\\
    \frac{\sqrt{2 m_d \omega_0} \lambda_3^{1/3}}{2 \times 10^{-3}} & n > 3,
\end{cases}
\end{equation}
where $10^{-3}$ is the typical DM velocity. 
In order to determine the appropriate phonon number $n$ for a given $\omega_{\rm th}$ we must take into account the subtlety that each excitation energy is smeared across a width, as discussed in Sec.~\ref{sec:toy_anharmonic} and also given in \eqref{eq:single_ph_width}. To solve for the smallest $n$ that contributes appreciably above $\omega_{\rm th}$, we solve the following equation:
\begin{equation} 
\label{eq:phonon_number}
\omega_{\mathrm{th}} = n \omega_0 + \sqrt{n} \sigma,
\end{equation}
where $\sigma$ is the single-phonon width as defined in \eqref{eq:single_ph_width}  and we have for simplicity taken $f(n) = n.$

Applying \eqref{eq:dm_mass_threshold}-\eqref{eq:phonon_number} to Si with $\omega_0 = 31$ meV, $\sigma = 18$ meV, and $m_d = 26$ GeV, we find the following results
\begin{equation} 
m_\chi \lesssim 
\begin{cases}
    \label{eq:dm_mass_threshold_si}
    0.2 \ \mathrm{MeV} \frac{\lambda_3}{10^{-2}} & \omega_\mathrm{th} = 80 \ \mathrm{meV}\\
    2.0 \ \mathrm{MeV} \Big( \frac{\lambda_3}{10^{-2}}\Big)^{1/2} & \omega_\mathrm{th}  = 120 \ \mathrm{meV}\\
    4.5 \ \mathrm{MeV} \Big( \frac{\lambda_3}{10^{-2}}\Big)^{1/3} & \omega_\mathrm{th} \geq  160  \ \mathrm{meV}
\end{cases}
\end{equation}
Below these masses, anharmonic corrections become large. The last line applies for  thresholds above 160 meV which corresponds to $n\ge 4$, and these $n$-phonon terms all give the same condition on DM mass. 
Note that this is only a heuristic, which does not include for example the combinatorial pre-factors or cancellations in the perturbation theory calculation. Nonetheless, we do see the same qualitative features in the complete numerical result which is given in Fig.~\ref{fig:anharmonic_ratios}.

\renewcommand{\arraystretch}{1.2}
\begin{table}
\centering
\begin{tabular}{ |c||c|c|c|  }
 \hline
 \multicolumn{4}{|c|}{Materials} \\
 \hline
 & $\omega_0$ [meV] & $\sigma$ [meV] & $\sqrt{2 m_d \omega_0}$ [keV]\\
 \hline
 GaAs   &   16.9  & 9.5 &  48.8 \\
 Ge&  18.2   &  10.6  & 49.6 \\
 Si & 30.8 & 17.6 & 40.3 \\
 Diamond  & 109.6 & 35.8 & 49.7 \\
 $\mathrm{Al}_2 \mathrm{O}_3$ &  51.6  &  20.4 & 51.1 \\
 \hline
\end{tabular}
\caption{ 
\label{fig:material_properties} \textbf{Single phonon properties for various crystals.} Using these energy scales, for a given experimental threshold we can estimate the DM masses where anharmonic effects become large, \eqref{eq:anharmonic_q}-\eqref{eq:dm_mass_threshold_si}. For crystals with non-identical atoms in a unit cell, we show the quantities averaged across atoms. The relative importance of anharmonic effects in the different materials will mainly be governed by the different phonon energies $\omega_0$. }
\end{table}

In order to generalize \eqref{eq:dm_mass_threshold_si} to other materials, we give the necessary energy scales in Tab.~\ref{fig:material_properties}. Despite large differences in $\omega_0$, the momentum scale $\sqrt{2 m_d \omega_0}$ ends up being about the same in all crystals. Then the typical DM mass scale for anharmonic effects to become important is also about the same for a fixed phonon number $n$. However, the differences in $\omega_0$ mean that the threshold corresponding to a given $n$ can vary significantly. For a given threshold, GaAs and Ge have the largest phonon number. Since anharmonic corrections become more important with larger $n$, GaAs and Ge will therefore have larger anharmonic contributions compared to Diamond at the same threshold.

In Fig.~\ref{fig:anharmonic_ratios}, we present the ratio of scattering rates in the anharmonic case to the harmonic case in Si and Ge, taking two representative cases for the couplings. We also present the cross-sections corresponding to an observed rate of 3 events per kg-yr in Fig.~\ref{fig:cross_sections}. The bands depict the possible uncertainty that anharmonicity introduces to an experimental reach, with the solid line giving the harmonic result and the dot-dashed the result for maximal anharmonicity. We do not show the effects above the cross sections of $\sigma_n \gtrsim 10^{-28}~\text{cm}^2$ as for these large interaction strengths, the DM is expected to lose a significant energy in 1 km of Earth's crust through scattering, thus rendering DM with such cross sections unobservable in underground direct detection experiments~\cite{Emken_2019}.

For $m_\chi > 10$ MeV, the typical $q$ becomes similar or larger than $\sqrt{2m_d \omega_0}$, where there is negligible difference in the anharmonic and harmonic structure factors. The rates will also start to be dominated by the impulse regime $q \gg \sqrt{2m_d \omega_0}$. In this case, the structure factor calculated with an anharmonic potential is nearly identical to that calculated in the harmonic case, as discussed in Sec.~\ref{sec:impulse}. We have also seen this behavior with numerical computations in Fig.~\ref{fig:impulse_approx}. The anharmonic and harmonic scattering rates are also essentially identical for DM masses $m_\chi > 10$ MeV.

For DM masses $m_\chi < 10$ MeV (i.e. $q < \sqrt{2m_d\omega_0}$), the ratio of the anharmonic to harmonic rate begins to grow with decreasing DM mass. As the typical $q$ decreases with decreasing DM mass, the leading anharmonic term~$\propto \frac{q^2}{2m_d\omega_0}$ grows faster compared to the harmonic term $\propto \Big(\frac{q^2}{2m_d\omega_0}\Big)^n$ for $n \geq 2$. The effect is more pronounced for higher thresholds or equivalently higher $n$, since the harmonic term is even more suppressed. Therefore at larger thresholds, the anharmonic effects start becoming important already at larger masses and also grows much more quickly as the DM mass is decreased. For a given DM mass, this also implies that the spectrum of events will have larger anharmonic corrections on the high energy tail of events. However, the rates are also highly suppressed in this tail, and only observable for high scattering cross sections.

At DM masses $m_\chi < 1$ MeV, the slope of the ratio of the anharmonic rate to the harmonic rate starts to decrease slightly, which is an artifact of the Brillouin zone momentum cutoff that we apply across all rate calculations. The incoherent and subsequent approximations are not guaranteed to be justified in this regime, so this effect should not be treated as physical. For sub-MeV DM masses, the phonons again should be treated as collective excitations, similar to the calculation of Ref.~\cite{Campbell-Deem:2019hdx}. 

Lastly, we note an interesting feature that the anharmonic scattering rate is strictly greater than the harmonic rate in the entire parameter space that we probe. This is a consequence of the sign of the leading $q$-scaling term $\frac{q^2}{2m_d \omega_0}$. For the production of an excited state $|\Phi_f\rangle$ in the crystal, the term in the dynamic structure factor $\propto q^2$ can only come from the term $|\langle \Phi_f|iqx|\Phi_0\rangle|^2$, as the mixing term $\propto \langle \Phi_f|I|\Phi_0\rangle \langle \Phi_f|\frac{(iqx)^2}{2}|\Phi_0\rangle^*$ and its conjugate are zero from orthogonality. Thus, the sign of the term $\propto q^2$ in the anharmonic structure factor is strictly positive for producing an excited state, whereas there is no corresponding term $\propto q^2$ in the harmonic case for $n \geq 2$ phonons. Since we are probing the $q \ll \sqrt{2 m_d \omega_0}$ regime, this leading term quickly dominates the structure factor. Thus, the anharmonic scattering rate exceeds the harmonic rate in this regime. A consequence of this is that we expect the harmonic crystal result gives a lower bound on the scattering.

\section{Conclusions}\label{sec:conclusions}

Scattering of DM with nuclei in crystals necessarily goes through production of one or many phonons for DM masses smaller than $\sim 100$ MeV. Previous work has focused on calculating the multiphonon scattering rates in a harmonic crystal under the incoherent approximation (i.e. $ q> q_{\mathrm{BZ}}$ or DM mass $\gtrsim$ MeV). In this work, we have studied the effects of anharmonicities in the crystal on the scattering rates, while still working within the incoherent approximation. 

In order to obtain a tractable calculation of anharmonic effects, we have simplified the problem into a toy model of a single atom in a 1D anharmonic potential. In this toy model, scattering into multiphonons can still be well-approximated by applying a smearing on the spectrum of quantized states to account for the phonon spectrum of a lattice. We extract anharmonic couplings by modeling the interatomic potentials of Si and Ge, which give rise to realistic single atom potentials. This approach allows us to obtain an analytic understanding and first estimate of the impact of anharmonicity, although the numerical results should not be taken as a definitive rate calculation.

We find that the harmonic crystal results of Ref.~\cite{Campbell-Deem:2022fqm} can be safely assumed for DM masses down to $\sim 10$ MeV. Below $\sim 10$ MeV, this assumption cannot be taken for granted. In this regime, we find that anharmonic effects on the scattering rates increase with decreasing DM mass and increasing experimental thresholds. Anharmonic corrections up to two orders of magnitude are possible for DM masses $\sim$ a few MeV and for experimental thresholds $\sim$ a few times the typical single phonon energy of the crystal. These findings are consistent with Refs.~\cite{Campbell-Deem:2019hdx,Campbell-Deem:2022fqm}, which studied two-phonon production from sub-MeV DM and found up to an order of magnitude larger rate from anharmonic couplings.

The size of the corrections is dependent on the material through the anharmonicity strength of that crystal and also, non-trivially, through the typical single phonon energies of the material. For a particular energy threshold, crystals with lower single phonon energies exhibit larger corrections since they require larger phonon numbers to be produced. For example, anharmonic effects in Ge can be larger by almost an order of magnitude than those in Si for similar DM parameter space and thresholds, even though the anharmonic couplings in the two crystals are similar. This is a consequence of the difference in $q$ scaling of the harmonic and anharmonic contributions, which become more pronounced with larger phonon number. Materials with low single-phonon energies, such as GaAs and Ge, therefore have the largest anharmonic effects. The effects will be reduced in Diamond and Al$_2$O$_3$, which have even higher single phonon energies than Si.

The relevance of anharmonic effects to direct detection experiments depends on the DM cross section. The effects are largest for low DM masses and high thresholds, in other words on the tails of the recoil spectrum where the rates are small. For a typical benchmark exposure of 1 kg-yr, the anharmonic corrections become sizeable for DM-nucleon cross sections above $\sim 10^{-34}$~cm$^2$. Being agnostic about any terrestrial or astrophysical constraints on the DM model and only requiring the DM to be observable in underground direct detection experiments, the upper bound on the DM cross section is $\sigma_n \lesssim 10^{-28}~\text{cm}^2$~\cite{Emken_2019}. This comes from considering an overburden of $\sim$ km. On the other hand, these very high DM-nucleon cross sections are typically excluded by terrestrial and astrophysical constraints for the simplest sub-GeV dark matter models~\cite{Knapen:2017xzo,Green:2017ybv}. DM-nucleon cross sections $\sigma_n \gtrsim 10^{-41}~\text{cm}^2$ ($\sigma_n \gtrsim 10^{-31}~\text{cm}^2$) are constrained for typical models with a heavy mediator (light dark photon mediator) for a DM mass $\sim$ MeV. With these constraints, we see from Fig.~\ref{fig:cross_sections} that the anharmonic effects can only impart corrections of at most an order of magnitude for experiments with kg-yr exposure. 

Experiments with exposures above kg-yr could see larger anharmonic effects, since they would be more sensitive to the events at high phonon number for MeV-scale DM. However, for solid-state direct detection experiments, achieving exposures significantly bigger than a kg-yr is challenging. Thus, for near-future crystal target experiments, we conclude that the anharmonic effects are only important up to $\mathcal{O}(1)$ factors at masses of $\sim$ a few MeV for the simplest DM models.

\acknowledgments
We are grateful to Simon Knapen and Xiaochuan Lu for useful discussions, and Simon Knapen for feedback on the draft. TL and EV were supported by Department of Energy grant DE-SC0022104. EV was also supported by a Sloan Scholar Fellowship. MS and CHS were supported by Department of Energy Grants DE-SC0009919 and DE-SC0022104. CHS was also supported by the Ministry of Education, Taiwan (MOE Yushan Young Scholar grant NTU-112V1039).

\appendix

\section{Interatomic potentials}\label{appendix:interatomic_potentials}

In order to produce results for a real crystal, we adopt atomic potentials based on Ref.~\cite{Rohskopf:2017}. The interatomic potentials used here are a combination of various commonly used empirical potentials. We choose to use the Tersoff-Buckingham-Coulomb interatomic potential defined in Ref.~\cite{Rohskopf:2017} using the parameters in the set labeled ``TBC-1", though other interatomic potentials may be chosen and give similar estimates for the anharmonicity strengths.

This potential includes a three-body Tersoff potential, originally defined in \cite{PhysRevB.37.6991}, which we restate here for reference.
\begin{align}
    \nonumber
    E &= \frac{1}{2} \sum_i \sum_{i \neq j} V_{ij} \\
    V_{ij} & = f_C (r_{ij}) \big(f_R (r_{ij}) + b_{ij} f_A (r_{ij}) \big),
    \label{eq:tersoff}
\end{align}
where the sum is over nearest-neighbor, and $r_{ij}$ is the distance between neighbors $i, j$. The function $f_C$ is a cutoff function that keeps the interaction short ranged, $f_R$ and $f_A$ are repulsive and attractive interactions, and $b_{ij}$ is a three-body term that is a function of the bonding angle of the third body with the atoms $i,j$. Explicitly, these functions are defined as
\begin{align}
& f_C(r)=\left\{\begin{array}{cll}
1 & \ \ r<R-D \\
\frac{1}{2}-\frac{1}{2} \sin \left(\frac{\pi}{2} \frac{r-R}{D}\right) & \ \ R-D<r<R+D \\
0 & \ \ r>R+D
\end{array}\right.
\\
& f_R(r)=A \exp \left(-\lambda_1 r\right) \\
& f_A(r)=-B \exp \left(-\lambda_2 r\right) \\
& b_{i j}=\left(1+\beta^n \zeta_{i j}{ }^n\right)^{-\frac{1}{2 n}} \\
& \zeta_{i j}=\sum_{k \neq i, j} f_C\left(r_{i k} \right) g\left[\theta_{i j k}\left(r_{i j}, r_{i k}\right)\right]
\nonumber
\\
&  \qquad \quad  \times \exp \left[\lambda_3{ }^m\left(r_{i j}-r_{i k}\right)^m\right] \\
& g(\theta)= 1+\frac{c^2}{d^2}-\frac{c^2}{\left[d^2+\left(\cos \theta-\cos \theta_0\right)^2\right]},
\end{align}
where $\theta_{ijk}$ is the angle between the displacement vectors $r_{ij}$ and $r_{ik}$. $R, D, A,B, \beta, n, c, d, \theta_0, \lambda_1, \lambda_2, \lambda_3$ are constants that can be found in Ref.~\cite{Rohskopf:2017}. Note that the notation in this section matches that of Ref.~\cite{Rohskopf:2017} and is standalone from the main text. Specifically, the parameters $\lambda_1, \lambda_2, \lambda_3$ are not to be confused with the anharmonicity strengths defined in the main text. In practice, anharmonicity arises from the asymmetry between the repulsive and attractive terms. The directional dependence of the anharmonicity strength is a result of the crystal's zincblende structure and bond angle-dependent potential.

The other components of this interatomic model include a long-range two-body Buckingham term
\begin{equation}
    V(r) = C e^{- r/\rho} - \frac{E}{r^6},
\end{equation}
and a screened Coulombic interaction defined by
\begin{align}
\nonumber
V(r) &= q^2 \Big[ \frac{\mathrm{erfc}(\alpha r)}{r} - \frac{\mathrm{erfc}(\alpha r_c)}{r_c} \\ 
& + \Big( \frac{\mathrm{erfc}(\alpha r_c)}{r_c^2} + \frac{2\alpha}{\sqrt{\pi}} \frac{e^{-\alpha^2 r_c^2}}{r_c}  \Big)(r-r_c) \Big] \\
& \times \Theta(r_c - r)
\end{align}
Here $q$ is the effective atomic charge, $\alpha$ is a damping parameter, and $r_c$ is a cutoff. As discussed in Ref.~\cite{Rohskopf:2017}, the full interatomic potential model is a sum of the three aforementioned interactions. All of the free parameters are fit onto the actual second, third, and fourth order forces calculated from DFT. This gives an analytic interatomic potential that produces the correct single-phonon dispersions and also captures the anharmonicity in the potential by fitting onto the higher order interatomic forces from DFT.

\section{Power counting in perturbation theory}\label{appendix:perturbation_theory}
In this appendix, we work out the explicit relation between the powers of $q^2$ and $\lambda_k$ in the perturbation theory calculation for the anharmonic Hamiltonian in \eqref{eq:toy_hamiltonian_perturbation}. 

The primary object we focus on in the dynamic structure factor is the squared matrix element $|\langle \Phi_n|e^{iqx}|\Phi_0\rangle |^2$, where $|\Phi_n \rangle$ are the eigenstates of the anharmonic Hamiltonian. With perturbation theory, the eigenstates can be expanded in powers of $\lambda_k$ as in \eqref{eq:perturbedeigenstate}. The corrections to the $n$th final state up to second order in $\lambda_k$ are given by,
\begin{align}
\label{eq:correction_first_order}
|\psi_n^{(1)}\rangle &= \sum_{k \neq n} \frac{V_{kn}}{(n-k)} |k\rangle ,\\
|\psi_n^{(2)}\rangle &= \sum_{k \neq n}\sum_{l \neq n} \frac{V_{kl}V_{ln}}{(n-k)(n-l)} |k\rangle -\frac{1}{2} |n\rangle \sum_{k \neq n}\frac{|V_{kn}|^2}{(n-k)^2} , \nonumber
\end{align}
where $V_{ij}\equiv \mel{i}{(\sqrt{2m_d\omega_0}x)^k}{j}$. In terms of the standard ladder operators of the harmonic oscillator, $V_{ij}$ are given by,
\begin{align}
V_{ij} = \mel{i}{(a+a^{\dagger})^k}{j}.
\end{align}
This tells us that $V_{ij}$ can only be non-zero when $i-j$ is one of the following: $-k$, $-k+2$,..., $k-2$, $k$. 

With these selection rules, the corrections in Eqs.~\ref{eq:correction_first_order} can be schematically written as,
\begin{align}
|\psi_n^{(1)}\rangle &\sim |n-k\rangle + |n-k+2\rangle + ... \nonumber \\& \quad + |n+k-2\rangle + |n+k\rangle \\
|\psi_n^{(2)}\rangle &\sim |n-2k\rangle + |n-2k+2\rangle + ... \nonumber \\& \quad + |n+2k-2\rangle + |n+2k\rangle
\end{align}
This pattern continues for higher orders in $\lambda_k$ such that at $\mathcal{O}(\lambda_k^j)$, we have,
\begin{align}\label{eq:perturbed_eigenstate_expansion}
|\psi_n^{(j)}\rangle &\sim |n-(j\times k)\rangle + |n-(j\times k)+2\rangle + ... \nonumber \\& \quad + |n+(j\times k)-2\rangle + |n+(j\times k)\rangle.
\end{align}
Note that the sum should only include terms for which the integer labelling the state is non-negative. With the knowledge of the unperturbed states appearing in $|\Phi_n\rangle$, the matrix element $\langle \Phi_n | e^{iqx}|\Phi_0\rangle$ can also be expanded in $\lambda_k$,
\begin{align}
\langle \Phi_n | e^{iqx}|\Phi_0\rangle \sim b_0 + \lambda_k b_1 + \lambda_k^2 b_2 + ...,
\end{align}
where the coefficients $b_j$ are given by,
\begin{align}\label{eq:bcoefficients}
b_0 &\sim \langle n | e^{iqx}|0\rangle \nonumber \\
b_1 &\sim \langle \psi_n^{(1)} | e^{iqx}|0\rangle + \langle n | e^{iqx}|\psi_0^{(1)}\rangle \\
b_2 &\sim \langle \psi_n^{(2)} | e^{iqx}|0\rangle + \langle \psi_n^{(1)} | e^{iqx}|\psi_0^{(1)}\rangle + \langle n | e^{iqx}|\psi_0^{(2)}\rangle \nonumber
\end{align}
In general, the coefficient $b_j$ is schematically given by,
\begin{align}\label{eq:b_structure}
b_j &\sim \langle \psi_n^{(j)} | e^{iqx}|0\rangle + \langle \psi_n^{(j-1)} | e^{iqx}|\psi_0^{(1)}\rangle + ... \nonumber \\&+ \langle \psi_n^{(1)} | e^{iqx}|\psi_0^{(j-1)}\rangle + \langle n | e^{iqx}|\psi_0^{(j)}\rangle.
\end{align}

To study the powers of $q$ appearing in $b_j$, we first need to understand the structure of the matrix element $\langle n_1 |e^{iqx}|n_2 \rangle$ for general eigenstates $|n_1\rangle$ and $|n_2\rangle$ of the unperturbed harmonic oscillator. This matrix element is given by the following,
\begin{align}
\langle n_1 |e^{iqx}|n_2 \rangle &= \sum_{l= \frac{n_1-n_2 + \vert n_1 - n_2 \vert}{2}}^{n_1}\frac{\sqrt{n_1! n_2!}}{l! (n_1 - l)! (n_2 - n_1 + l)!} \times \nonumber\\&\quad \quad \Big(\frac{iq}{\sqrt{2m_d \omega_0}} \Big)^{n_2 - n_1 + 2l} e^{- \frac{q^2}{4 m_d \omega_0}}.
\end{align}
We learn that the matrix element $\langle n_1 |e^{iqx}|n_2 \rangle $ contains powers of $iq/(\sqrt{2m_d\omega_0})$ ranging from $|n_1 - n_2|$ to $n_1 + n_2$. Note again that the Debye-Waller factor $e^{- \frac{q^2}{4 m_d \omega_0}}$ is not included in this power counting since $e^{- \frac{q^2}{4 m_d \omega_0}} \approx 1$ in the regime of interest.

Combining this information with the structure of $b_j$ in \eqref{eq:b_structure} and the structure of $|\psi_n^{(j)}\rangle$ in \eqref{eq:perturbed_eigenstate_expansion}, the powers of $q$ in $b_j$ can be identified:
\begin{align}
b_j \sim e^{- \frac{q^2}{4 m_d \omega_0}} &\Big\{\Big(\frac{iq}{\sqrt{2m_d\omega_0}}\Big)^{n-jk}+\Big(\frac{iq}{\sqrt{2m_d\omega_0}}\Big)^{n-jk+2} \nonumber \\ & \quad \quad +...+\Big(\frac{iq}{\sqrt{2m_d\omega_0}}\Big)^{n+ j k }\Big\}.
\end{align}
Note that only those terms with powers of $q$ larger or equal to 1 are present. Terms $\propto q^0$ have to cancel as they otherwise lead to $q^0$ terms in the squared matrix element $|\langle \Phi_n | e^{iqx}|\Phi_0\rangle|^2$, which is forbidden due to orthogonality of eigenstates. 

As the kinematic regime under consideration is of $q \ll \sqrt{2m_d\omega_0}$, we will focus on powers of $q$ less than $n$, which corresponds to the harmonic case. We see from the equation above that the lowest powers of $q$ decrease with increasing values of $j$. Thus, higher order corrections in $\lambda_k$ appear with lower powers in $q$. Eventually, at a sufficiently high power of $\lambda_k$, we get a coefficient $b_j$ with the minimum power of $q$ equal to 1. The squared matrix element can then be written in general as,
\begin{align}\label{eq:squared_matrix_element_appendix}
&|\langle \Phi_n | e^{iqx}|\Phi_0\rangle|^2 = e^{- \frac{q^2}{2 m_d \omega_0} } \times \Bigg[\frac{1}{n!}\Big(\frac{q^2}{2m_d\omega_0}\Big)^n \nonumber \\&+ \sum_{i \geq 1} \Big(\frac{q^{2}}{2m_d\omega_0}\Big)^{i}   \left( a_{n,i}\, \lambda_k^{\nu(n,i)} + \mathcal{O}\Big(\lambda_k^{\nu(n,i)+1}\Big)\right)\Bigg], 
\end{align}
where the first term on the right hand side $\propto q^{2n}$ is the harmonic term, and the anharmonic corrections are expanded in powers of $q^2$ which are denoted by $i$, with $i \geq 1$. Every power $i$ appears with a minimum \textit{allowed} power $\nu(n,i)$ of $\lambda_k$. 

To study the behavior of $\nu(n,i)$, we first note that, for even $k$, the matrix element $\langle \Phi_n | e^{iqx}|\Phi_0\rangle$ is purely real or purely imaginary, depending on whether $n$ is even or odd respectively. For instance, if $n$ is even, then $b_0$ is purely real. Higher orders in $\lambda_k$ lead to insertions of $(a + a^\dagger)^k$ and therefore matrix elements where the difference in the harmonic oscillator states is also even, so that all coefficients $b_j$ are real in this case. But for odd $k$, the $b_j$ coefficients will alternate in being real and imaginary. This changes the structure of the squared matrix element depending on $k$, as we will see below. 

{\bf Odd $k$:} We will first consider odd $k$. In this case, the squared matrix element can be written as,
\begin{align}
|\langle \Phi_n |& e^{iqx}|\Phi_0\rangle|^2 \sim |b_0 + \lambda_k^2 b_2 + \lambda_k^4 b_4 + ...|^2 \nonumber \\
& \qquad \qquad \qquad + | \lambda_k b_1 + \lambda_k^3 b_3 + ...|^2 \\
& \sim |b_0|^2 + \lambda_k^2(|b_1|^2 + (b_0 b_2^* + b_0^* b_2)) \nonumber\\
&\qquad +\lambda_k^4 (|b_2|^2+(b_0 b_4^* + b_0^* b_4)+(b_1 b_3^* + b_1^* b_3)
)\nonumber\\
&\qquad +\mathcal{O}(\lambda_k^6) \\
&\sim e^{- \frac{q^2}{2 m_d \omega_0}}\Big[ \frac{1}{n!}\Big(\frac{q^2}{2m_d\omega_0}\Big)^n \nonumber\\
& \qquad + \lambda_k^2 \Big\{\Big(\frac{q^2}{2m_d\omega_0}\Big)^{n-k}+\Big(\frac{q^2}{2m_d\omega_0}\Big)^{n-k+1} \nonumber \\
& \qquad \quad + ... + \Big(\frac{q^2}{2m_d\omega_0}\Big)^{n+k}\Big\}\nonumber\\
& \qquad + \lambda_k^4 \Big\{\Big(\frac{q^2}{2m_d\omega_0}\Big)^{n-2k}+\Big(\frac{q^2}{2m_d\omega_0}\Big)^{n-2k+1} \nonumber \\
& \qquad \quad + ... + \Big(\frac{q^2}{2m_d\omega_0}\Big)^{n+2k}\Big\} + \mathcal{O}(\lambda_k^6)\Big].
\end{align}
Thus we see that we get corrections at even orders in $\lambda_k$, with the lowest non-zero power being $\lambda_k^2$. In general, at $\mathcal{O}(\lambda_k^j)$ for an even $j=2j'$, the lowest power of $q^2$ is $n-(j' \times k)$, and the highest power is $n+(j' \times k)$. Note that only terms with positive powers of $q^2$ are present. The term $\propto q^2$ can also subtly cancel in some cases as there is no term $\propto q^0$ in coefficients $b_j$. We will deal with this case later below. But to get a power $i>1$ of $q^2$, the lowest non-zero $j'$ is $\lceil\frac{|n-i|}{k}\rceil$, with the lowest $j$ given by $2 \times \lceil\frac{|n-i|}{k}\rceil$. Thus, in the squared matrix element, the lowest non-zero power $\nu(n,i)$ required is given by,
\begin{align}
\nu(n,i) = \text{max}\Big(2\times\lceil\frac{|n-i|}{k}\rceil~,~2\Big).
\end{align}

To get the lowest power $i=1$ of $q^2$ i.e. the term $\propto q^2$, the only possible way is to get the term $\propto q^1$ in the coefficient $b_j$ as there is no term $\propto q^0$. For odd $n$, the term $\propto q^1$ in $b_j$ can only be generated at an even $j$, since that is the only way to satisfy $n - jk = 1$. For every even $j=2j'$, the powers of $q$ in $b_j$ range from $n-(2k)\times j'$ to $n+(2k)\times j'$. The lowest $j'$  to get a term $\propto q^1$ is then given by $\lceil \frac{|n-1|}{2k}\rceil$, with $j$ given by $2 \times\lceil \frac{|n-1|}{2k}\rceil$. For an even $n$, the term $\propto q^1$ in $b_j$ can only be generated for an odd $j$. For every odd $j=2j'-1$, the lowest power of $q$ in $b_j$ is $n+k-(2k)\times j'$. The lowest $j'$  to get a term $\propto q^1$ is then given by $\lceil \frac{|n+k-1|}{2k}\rceil$, with $j$ given by $2 \times\lceil \frac{|n+k-1|}{2k}\rceil -1$. In the squared matrix element, the lowest non-zero power $\nu(n,1)$ required is given by,
\begin{equation} 
\nu(n,1) = 
\begin{cases}
    \text{max}\Big(4 \times \lceil \frac{|n-1|}{2k} \rceil~,~2 \Big) & \text{for odd} ~n\\
    4 \times \lceil \frac{|n+k-1|}{2k} \rceil - 2 & \text{for even} ~n
\end{cases}
\end{equation}

{\bf Even $k$:} Now we consider even $k$. In this case, the squared matrix element is,
\begin{align}
|\langle \Phi_n &| e^{iqx}|\Phi_0\rangle|^2 \sim |b_0 + \lambda_k b_1 + \lambda_k^2 b_2 + ...|^2  \\
& \sim |b_0|^2 + \lambda_k( (b_0 b_1^* + b_0^* b_1)) \nonumber\\
& \qquad +\lambda_k^2 (|b_1|^2+(b_0 b_2^* + b_0^* b_2)+\mathcal{O}(\lambda_k^3) \\
& \sim e^{- \frac{q^2}{2 m_d \omega_0}}\Big[ \frac{1}{n!} \Big(\frac{q^2}{2m_d\omega_0}\Big)^n \nonumber\\
& \qquad  + \lambda_k \Big\{\Big(\frac{q^2}{2m_d\omega_0}\Big)^{n-k/2}+\Big(\frac{q^2}{2m_d\omega_0}\Big)^{n-k/2+1} \nonumber \\
& \qquad \quad + ... + \Big(\frac{q^2}{2m_d\omega_0}\Big)^{n+k/2}\Big\}\nonumber\\
& \qquad + \lambda_k^2 \Big\{\Big(\frac{q^2}{2m_d\omega_0}\Big)^{n-k}+\Big(\frac{q^2}{2m_d\omega_0}\Big)^{n-k+1} \nonumber \\
& \qquad \quad  + ... + \Big(\frac{q^2}{2m_d\omega_0}\Big)^{n+k}\Big\} + \mathcal{O}(\lambda_k^3)\Big].
\end{align}
Thus we see that we get corrections at all orders in $\lambda_k$, with the lowest non-zero power being $\lambda_k$. In general, at $\mathcal{O}(\lambda_k^j)$, the lowest power of $q^2$ is $n-(j \times k)/2$, and the highest power is $n+(j \times k)/2$. Following similar arguments to the case of odd $k$ discussed earlier, $\nu(n,i)$ for $i>1$ is given by,
\begin{align}
\nu(n,i) = \text{max}\Big(\lceil\frac{|n-i|}{k/2}\rceil~,~1\Big).
\label{eq:evenk_nu}
\end{align}

Another difference between the case of even $k$ considered here and that of odd $k$ is that we do not get an $i=1$ term for even $n$, as all terms in the coefficients $b_j$ contain even powers of $q$. This means that the leading term will always go as $q^4$, with a $\lambda_k$ power determined by \eqref{eq:evenk_nu} for $i=2$. For odd $n$, the lowest power of $q$ in $b_j$ is $n-k \times j$. Thus, in the squared matrix element, the lowest non-zero power $\nu(n,1)$ required is given by,
\begin{align}
\nu(n,1) =  \text{max}\Big(2 \times \lceil \frac{|n-1|}{k}\rceil~,~1 \Big).
\end{align}

The calculations in this appendix up to this point consider the overall scaling behavior of the powers of $q^2$ and $\lambda_k$ in the squared matrix element. We have neglected combinatorial factors at several steps in the calculations that enter into the numerical coefficients $a_{n,i}$ in \eqref{eq:squared_matrix_element_appendix}. Sometimes, the numerical coefficients can also cancel with each other, and the naive leading behavior estimated in this section can vanish. In order to give concrete examples of the numerical coefficients, we perform explicit calculations of the squared matrix element using perturbation theory with $k=3$ (i.e. a cubic perturbation), and phonon numbers $n=1,~2,~3,~\text{and}~4$. We perform this explicit calculation only up to $\mathcal{O}(\lambda_3^2)$. The results of various numerical coefficients are presented below. 

For a single-phonon production (i.e. $n=1$), the coefficients $a_{n,i}$ are given by,
\begin{align}
a_{1,1}&=44\\
a_{1,2}&=-82\\
a_{1,3}&=5.
\end{align}
For a two-phonon production (i.e. $n=2$), the coefficients are given by,
\begin{align}
a_{2,1}&=8\\
a_{2,2}&=59\\
a_{2,3}&=-56\\
a_{2,4}&=2.5.
\end{align}
For a three-phonon production (i.e. $n=3$), the coefficients are,
\begin{align}
a_{3,2}&=18\\
a_{3,3}&=37\\
a_{3,4}&=-23.04\\
a_{3,5}&=0.77.
\end{align}
Note that we do not show the coefficient $a_{3,1}$ as it appears at $\mathcal{O}(\lambda_3^4)$. Finally, for a four-phonon production (i.e. $n=4$), the coefficients are evaluated to be,
\begin{align}
a_{4,1}&=0\\
a_{4,2}&=0\\
a_{4,3}&=0.097\\
a_{4,4}&=0.05\\
a_{4,5}&=-0.012\\ 
a_{4,6}&=1.81 \times 10^{-4}.
\end{align}
Note that the coefficients $a_{4,1}$ and $a_{4,2}$ amount to zero because of a numerical cancellation between the two terms in the $b_1$ coefficient in Eq.~\ref{eq:bcoefficients}. The leading behavior of the terms proportional to $q^2$ and $q^4$ in the structure factor is instead $q^2 \lambda_3^6$ and $q^4 \lambda_3^4$, respectively.

As these numerical coefficients appear through combinations and interferences of several combinatorial factors at various steps of the calculation, it is hard to provide a general expression for them. By looking at the examples above however, we can make some general observations. Typically, we see that the coefficients follow a pyramid structure, with $a_{n,i}$ being the largest for $i$ near $n$, and decreasing with $i$ away from $n$.  We also find that the coefficients can vary by orders of magnitude from each other. The terms with $i$ near $n$ receive contributions from several individual matrix elements, and in general seem to be larger. We expect to see this pattern continue for higher phonon numbers as well. The exact values of these coefficients play a role in determining where the anharmonic corrections dominate, and so our power counting approach only gives an $O(1)$ estimate.

\section{Impulse approximation}
\label{sec:impulse_appendix}

In Sec.~\ref{sec:impulse}, we calculated the structure factor via the saddle point approximation in the regime defined by \eqref{eq:impulse_regime}. This regime corresponded to values of $\omega$ near $\frac{q^2}{2m}$ and within the Gaussian width of \eqref{eq:impulse_result}. As discussed in the main text, in order to calculate the tail of the structure factor far from $\omega = \frac{q^2}{2m},$ more expansion terms are needed in $f$. Here we discuss this extension of the impulse approximation.

First, in the special case of a harmonic potential, we can start from the full result in Eq.~\eqref{eq:1dsqw_harmonic}. After rewriting the energy conservation delta function as a time integral, 
we find that
\begin{equation}
\label{eq:impulse_f_harmonic}
    f(t) = -i\omega t +\frac{q^2}{2m_d \omega_0} (e^{i\omega_0 t}-1)
\end{equation}
Solving $f'(t)=0$ gives the exact result
\begin{equation}
\label{eq:ti_harmonic_saddle_exact}
    t_I = \frac{i}{\omega_0}\ln \left(\frac{q^2}{2m_d\omega} \right).
\end{equation}
Using the saddle point approximation for $\omega \gg \omega_0$, we find
\begin{equation}
S_{\textrm{toy},d}(q,\omega) \sim \frac{1}{\sqrt{\omega \omega_0}}\,e^{-2W_{\rm toy}(q)} \left(\frac{q^2}{2m{\omega}}\right)^{\frac{\omega}{\omega_0}} e^{\frac{\omega}{\omega_0}}.
\label{eq:harmonic_saddle}
\end{equation}
The same result can also be derived by approximating the sum over phonon states as an integral in Eq.~\eqref{eq:1dsqw_harmonic}. The saddle point approximation for the harmonic oscillator holds as long as $\omega \gg \omega_0$, and we no longer have a condition on 
how close $\omega$ is to $\frac{q^2}{2m}$. In the impulse regime, $\omega \sim \frac{q^2}{2m}$, one can check that it reduces to the previous result in Eq.~\eqref{eq:impulse_result}. We see in this exact result that the tail at large $\omega$ is Poissonian instead of Gaussian.

For general potentials, this exact analytic result is no longer possible, but we can still calculate corrections to the tail. First, we start by giving the exact saddle point equation:
\begin{equation}
    0=f'(t_I) = - i \left (E_0 + \omega - \frac{q^2}{2m} \right)
    +i\frac{\left\langle H' e^{iH't_I}\right\rangle}{\left\langle e^{iH't_I}\right\rangle} 
    \label{eq:saddle_point_equation}
\end{equation}
which is valid at all orders. We begin by noticing that saddle point equation \eqref{eq:saddle_point_equation} is satisfied exactly at $\omega = \frac{q^2}{2m}$ by $t_I = 0$. Then, $\omega$-derivatives of $t_I$ at $\omega=\frac{q^2}{2m}$ can be found by taking $\omega$-derivatives of \eqref{eq:saddle_point_equation} and solving for $t_I^{(n)}[\omega = \frac{q^2}{2m}]$. This allows us to calculate $t_I[\omega=\frac{q^2}{2m}]$ in an iterative fashion. The first few terms are
\begin{align}
    \nonumber
    t_I[\frac{q^2}{2m}] &= 0
    \\ \nonumber
    t_I'[\frac{q^2}{2m}] &= \frac{i}{\langle H'\rangle^2 - \langle H'^2 \rangle}
    \\ \nonumber
    t_I''[\frac{q^2}{2m}] &= i\frac{-2 \langle H' \rangle^3 + 3 \langle H' \rangle \langle H'^2 \rangle - \langle H'^3 \rangle}{\big(\langle H'\rangle^2 - \langle H'^2 \rangle\big)^3}
    \\ \nonumber
    t_I^{(3)}[\frac{q^2}{2m}] &= \frac{i}{\big(\langle H'\rangle^2 - \langle H'^2 \rangle\big)^5} \times 
    \\ \nonumber
    & \Big( 6 \langle H' \rangle^6 - 18 \langle H' \rangle^4 \langle H'^2 \rangle + 3 \langle H'^2 \rangle^3
    \\ \label{eq:ti_expansion}
    & + 8 \langle H' \rangle^3 \langle H'^3 \rangle - 14 \langle H' \rangle \langle H'^2 \rangle \langle H'^3 \rangle \\ \nonumber
    & + 3 \langle H'^3 \rangle^2 - \langle H'^2 \rangle \langle H'^4 \rangle \\ \nonumber
    & + \langle H'^2 \rangle \big( 12 \langle H'^2 \rangle^2 + \langle H'^4 \rangle \big) \Big)
\end{align}
where $t_I^{(n)}$ denotes the $n$th $\omega$-derivative of $t_I$. In the harmonic case, this series resums to \eqref{eq:ti_harmonic_saddle_exact}. For general potentials, one can then use the expansions \eqref{eq:impulse_fourth_order} and \eqref{eq:ti_expansion} to calculate
\begin{equation}
    S_{\text{toy},d}(q,\omega) \approx
    \sqrt{ \frac{2 \pi}{-f''(t_I)}} e^{f(t_I)}
\end{equation}
to a desired order.

\section{Exact results for Morse potential}
\label{sec:morse}

\begin{figure}
\centering
\includegraphics[width=\linewidth]{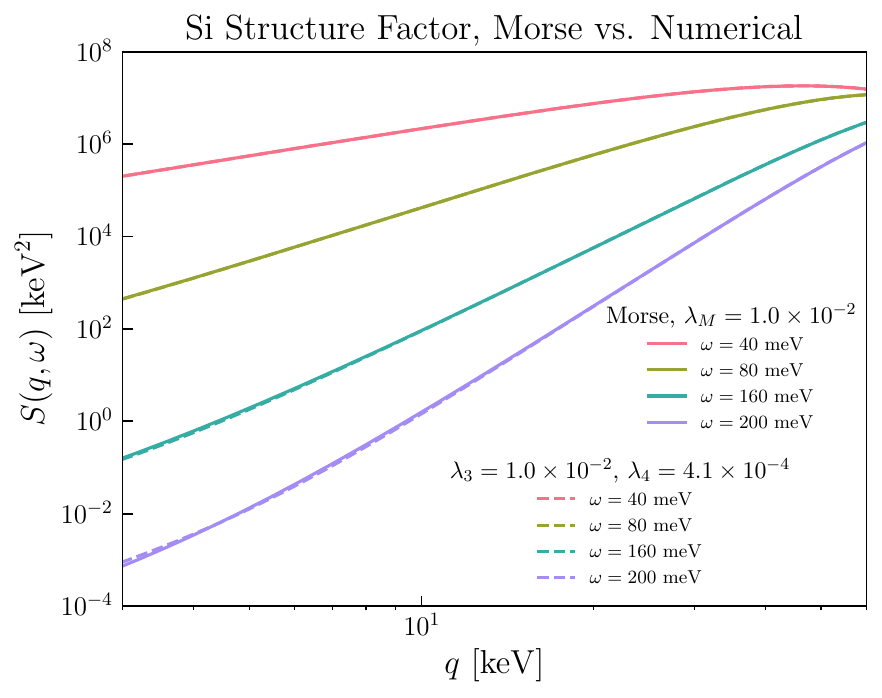}  
\caption{ 
\label{fig:structure_morse}  Comparison of analytic structure factor in the Morse potential and the numerical calculation for Si as described in Sec.~\ref{sec:exact}. We find that the two methods give almost the same result due to the fact that the Morse potential well approximates the single-atom potential along the nearest-neighbor direction.}
\end{figure}

\begin{figure*}
\centering
\includegraphics[width=0.98\linewidth]{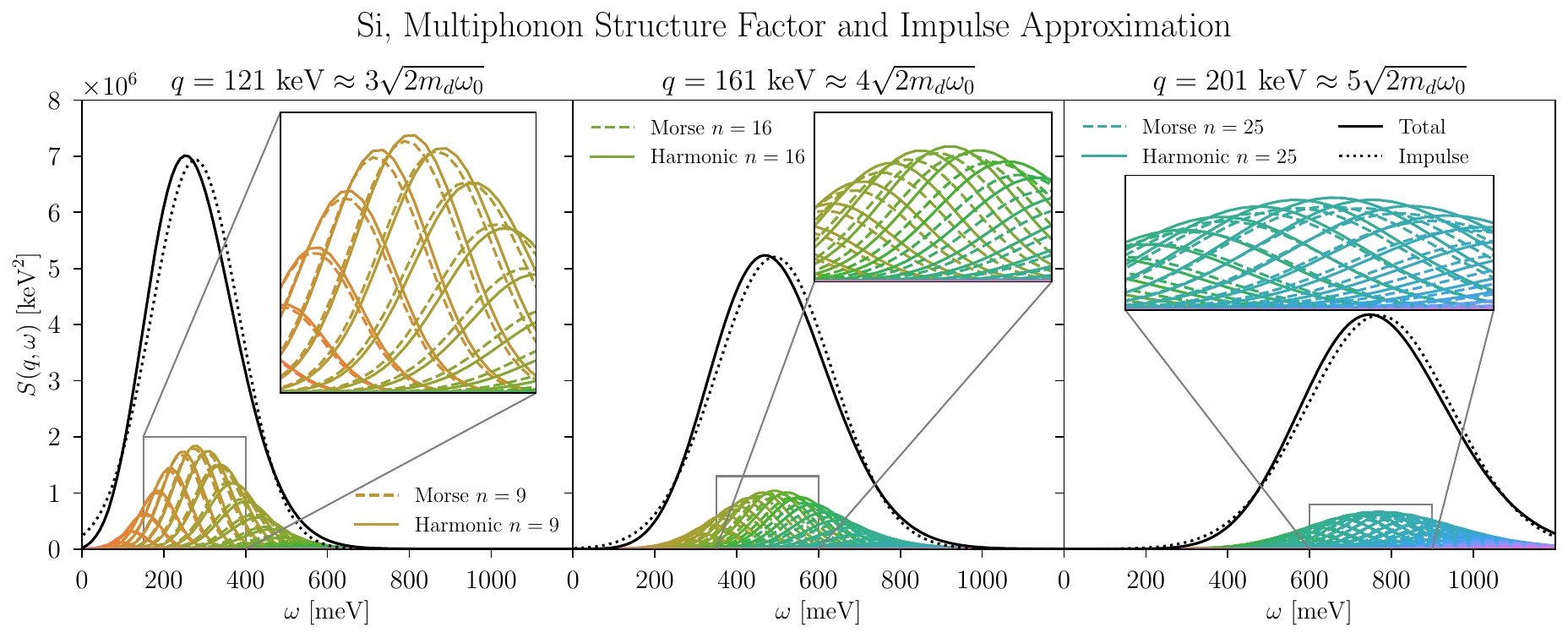}  
\caption{ 
\label{fig:impulse_morse} {\bf $\omega$-dependence of structure factor for the Morse potential:} Comparison of the Morse (dashed rainbow) and harmonic (solid rainbow) structure factor contributions from each individual excited state. The solid black line is the sum of contributions from the Morse potential, and dotted is the impulse envelope. Even though the energies of each individual Morse excited state are perturbed, the total structure factor remains essentially unchanged from the harmonic result. The small shift of order $\omega_0$ between the exact result and impulse approximation results from dropping higher order terms in the impulse approximation as discussed in Sec.~\ref{sec:impulse} and App.~\ref{sec:impulse_appendix}.}
\end{figure*}

The Morse potential is a special case of an anharmonic potential where the structure factor is analytically solvable. We will use this case to illustrate the behavior of the structure factor discussed in Sec.~\ref{sec:perturbation_theory}.  We also use it to validate the numerical calculations used in our final results and check the validity of the impulse approximation in the regime where there are $n>10$ phonons.

The Morse potential is defined as
\begin{equation}
    \label{eq:Morse_hamiltonian}
    V_{\mathrm{Morse}} =  B \Big( e^{-2 a x} - 2 e^{-a x} \Big),
\end{equation}
where $a$ is a parameter controlling the width of the potential and $B$ is the normalization. Expanding this potential in powers of $x$ gives
\begin{align}
V_{\mathrm{Morse}}= -B + Ba^2x^2 - B a^3 x^3 +\frac{7}{12} B a^4 x^4 + ...
\end{align}
Matching the quadratic and the cubic terms with \eqref{eq:toy_anharmonic_potential}, we find that
\begin{align}
a &= -4 \lambda_3 \sqrt{2 m \omega_0} \\ 
B &= \frac{\omega_0}{64 \lambda_3^2}.
\end{align}
Note that the Morse potential has fewer free parameters than the anharmonic potential up to fourth order in the displacements, so we cannot simultaneously fit $\lambda_4$. Nonetheless, the realistic potential as obtained App.~\ref{appendix:interatomic_potentials} are well approximated by this Morse potential due to the dominance and Morse-like behavior of the $f_R$ and $f_A$ terms in the Tersoff part of the potential.

The Morse potential approximation of our anharmonic potential is then given by 
\begin{equation}
    V_{\mathrm{Morse}} = \frac{\omega_0}{64 \lambda_M^2} \Big( e^{8 \lambda_M \sqrt{2 m \omega_0}} - 2 e^{4 \lambda_M \sqrt{2 m \omega_0} x} \Big),
\end{equation}
where we take $\lambda_M = \lambda_3$ in order to fit up to third order anharmonicities. In this potential, the structure factor \eqref{eq:structure_factor_final} is exactly calculable since the Morse eigenstates and eigenenergies are known analytically. These results~\cite{Berrondo_1987} give squared matrix elements between the ground state and $n$th excited state of
\begin{align}
    \nonumber
    & \vert \langle \Phi_n \vert e^{iqx} \vert \Phi_0 \rangle \vert^2 = \frac{(2K-2n-1)(2K-1)}{n!\Gamma(2K) \Gamma(2K-n)} \\ & \times \abs{\frac{\Gamma (n + \frac{i(q/\sqrt{2m \omega_0})}{4\lambda_M}) \Gamma(2K + \frac{i(q/\sqrt{2m \omega_0})}{4\lambda_M} - n - 1)}{\Gamma(\frac{iq/\sqrt{2m \omega_0}}{4\lambda_M})}}^2,
    \label{eq:morse_eiqx}
\end{align}
with energy gaps
\begin{equation}
    \label{eq:morse_energies}
    E_n - E_0 = \big( n -  \frac{n (1+n)}{2K} \Big) \omega_0,
\end{equation}
where $K = \frac{1}{32 \lambda_M^2}$. 

Note that these formulae are only valid for $n < K - \frac{1}{2}$ since above this excited state, the eigenstates are unbound and have a different analytic form. For $\lambda_M \sim 0.01$, this condition requires $n \lesssim 312$, which corresponds to an energy gap of $\mathcal{O}(\mathrm{eV})$. Recoil energies at this scale are comparable to the size of a typical lattice potential well and thus the free nuclear recoil approximation holds. Then, for typical anharmonicity strengths, the discrete states of the Morse potential that we have used in this analysis are sufficient to estimate the anharmonic effect in the multiphonon scattering regime.

Putting together expressions \eqref{eq:morse_eiqx}-\eqref{eq:p2_morse}, \eqref{eq:structure_factor_final}, and \eqref{eq:structure_factor_impulse} we can calculate the structure factor in the Morse potential in both the large and small $q$ regime. We show these results for $\lambda_M = \lambda_3$ in Figs.~\ref{fig:structure_morse} and \ref{fig:impulse_morse}. 
Fig.~\ref{fig:structure_morse} provides a check for our numerical results in Sec.~\ref{sec:exact}. Here we see that the numerical calculations and corresponding analytic Morse results are almost identical. There is a modified $q$ scaling of the structure factor compared to the harmonic case, as was already illustrated in Fig.~\ref{fig:si_structure_factor_q}. We can also obtain this behavior analytically with the Morse potential. Expanding the expression \eqref{eq:morse_eiqx} to leading order in $q$ and subsequently in $\lambda_M$, we get explicitly, 
\begin{align}
    \vert \langle 2 \vert e^{iqx} \vert 0 \rangle \vert^2 &= 8 \lambda_M^2 q^2 + \ldots
    \\
    \vert \langle 3 \vert e^{iqx} \vert 0 \rangle \vert^2 &= \frac{512}{3} \lambda_M^4 q^2 + \ldots
    \\
    \vert \langle 4 \vert e^{iqx} \vert 0 \rangle \vert^2 &= 6144 \lambda_M^6 q^2 + \ldots,
    \\
    \vert \langle 5 \vert e^{iqx} \vert 0 \rangle \vert^2 &= \frac{1572864}{5} \lambda_M^8 q^2 + \ldots,
    \\
    \vert \langle 6 \vert e^{iqx} \vert 0 \rangle \vert^2 &= 20971520 \lambda_M^{10} q^2 + \ldots,
\end{align}
where the ellipses include higher orders in both $q$ and $\lambda_M$. The leading $\lambda_M$ scalings are consistent with those illustrated in Fig.~\ref{fig:anharmonic_schematic} for $n=2$ and $3$. For $n=4$, the leading $\lambda_M$ scaling differs from the power counting in Fig.~\ref{fig:anharmonic_schematic}, but matches with the explicit results obtained using perturbation theory as presented in Appendix~\ref{appendix:perturbation_theory}. An exact numerical cancellation modifies the leading behavior to $\lambda_M^6 q^2$. We see that the leading behavior in $q, \lambda_M$ for $n>4$ also differs from the $x^3$-theory power counting, suggesting a generic presence of cancellations at lower orders of $\lambda_3$ for the $q^2$ dependence.

In Fig.~\ref{fig:impulse_morse}, we demonstrate that the impulse approximation remains robust for $q \gg \sqrt{2 m_d \omega_0}$ and $n>10$ excited states. Note that we can also calculate corrections to $\langle p^2 \rangle$ in the Morse ground state exactly:
\begin{equation}
    \langle p^2 \rangle = \frac{m \omega_0}{2} (1 - 16 \lambda_M^2),
    \label{eq:p2_morse}
\end{equation}
which is used in the impulse regime result \eqref{eq:impulse_result}.  The impulse result is almost identical between the Morse and harmonic cases, since the Gaussian width is only corrected at order $\lambda_M^2,$ which is $\sim 10^{-4}.$ This is also borne out in the full calculation of the structure factor shown in Fig.~\ref{fig:impulse_morse}.

\bibliographystyle{apsrev4-1}
\bibliography{phonons}
 
\end{document}